\documentclass[fleqn]{2023SCGE}
\usepackage{amsmath} 
\usepackage{amsthm} 
\newtheorem{mythm}{Theorem} 
\usepackage[numbers]{natbib}

\begin{document}

\ensubject{subject}

\ArticleType{Article}
\SpecialTopic{SPECIAL TOPIC: }
\Year{2023}
\Month{January}
\Vol{66}
\No{1}
\DOI{??}
\ArtNo{000000}
\ReceiveDate{January 11, 2023}
\AcceptDate{April 6, 2023}

%


\title{The pseudospectrum and spectrum (in)stability of quantum corrected Schwarzschild black hole}

\author[2,3]{Li-Ming Cao}{}%
\author[1]{Jia-Ning Chen}{}
\author[1]{Liang-Bi Wu}{{liangbi@mail.ustc.edu.cn}}
\author[1]{Libo Xie}{}
\author[3]{Yu-Sen Zhou}{}

%

\address[1]{School of Fundamental Physics and Mathematical Sciences, Hangzhou Institute for Advanced Study, UCAS, Hangzhou 310024, China}
\address[2]{Peng Huanwu Center for Fundamental Theory, Hefei, Anhui 230026, China}
\address[3]{Interdisciplinary Center for Theoretical Study and Department of Modern Physics,\\
	University of Science and Technology of China, Hefei, Anhui 230026,
	China}


\abstract{
	In this study, we investigate the pseudospectrum and spectrum (in)stability of quantum corrected Schwarzschild black hole. Methodologically, we use the hyperboloidal framework to cast the quasinormal mode (QNM) problem into an eigenvalue problem associated with a non-selfadjoint operator, and then the spectrum and pseudospectrum are depicted. Besides, the invariant subspace method is exploited to improve the computational efficiency for pseudospectrum. The investigation into the spectrum (in)stability entails
	two main aspects. On the one hand, we calculate the spectra of the quantum corrected black hole, then by the means of the migration ratio, the impact of the quantum correction effect on the Schwarzschild black hole has been studied. The results indicate that the so-called ``migration ratio instability" will occur for small black holes with small angular momentum number $l$. In the eikonal limit, the migration ratios remain the same for each overtone. On the other hand, we study the spectrum (in)stability of the quantum corrected black hole by directly adding some particular perturbations into the effective potential, where perturbations are located at the event horizon and null infinity, respectively. There are two interesting observations under the same perturbation energy norm. First, perturbations at infinity are more capable of generating spectrum instability than those at the event horizon. Second, we find that the peak distribution can lead to the instability of QNM spectrum more efficiently than the average distribution.}



\maketitle


\begin{multicols}{2}
\section{Introduction}
In a strong field environment, one of the most significant tests of general relativity (GR) is the detection of gravitational waves (GWs) from the merger of binary black holes and neutron stars at present~\cite{LIGOScientific:2016aoc,LIGOScientific:2016sjg,LIGOScientific:2017bnn,LIGOScientific:2017vwq,KAGRA:2021vkt}. The ringdown stage, occurring after the merger, consists of exponentially damped oscillatory signals that are superimposed by the so-called quasinormal modes (QNMs)~\cite{Berti:2009kk,Konoplya:2011qq,Franchini:2023eda}. \Authorfootnote QNMs provide valuable insights for us to understand black holes (BHs), because the characteristic frequencies of QNMs, which are complex since the perturbed BHs are the dissipative system, are related with the parameters of BHs.

In recent years, due to the fact that BHs are not isolated and they are always surrounded by some matters, the QNM spectrum of such BHs from Einstein gravity affected by the astrophysical environment has gained considerable attention as the improvement of the accuracy of the GWs detection~\cite{Barausse:2014tra}. The (in)stability of the QNM spectrum provides an analysis for the impact. Initial studies by Nollert and Price~\cite{Nollert:1996rf,Nollert:1998ys} demonstrated that small perturbations can
significantly alter the QNM spectrum, a phenomenon known as spectrum instability. What in their works tell us that small perturbations can involve significant changes on the QNM spectrum, which is called the spectrum instability. Besides, Daghigh\,\textit{et al.}\,find that even with a continuous piecewise linear approximation of the Regge-Wheeler potential, there is still a significant difference in the characteristic frequencies compared to the original potential~\cite{Daghigh:2020jyk}. Because astrophysical BHs are not isolated, but rather are permeated by environments, quantum corrections, or other generic modifications. Therefore, it is necessary to understand the effects of these factors on QNM spectrum and the (in)stability of QNM spectrum.

When the forms of external perturbations are known, the spectrum (in)stability can be studied naturally. However, without knowing the specfic form of external perturbation, a new powerful approach, pseudospectrum, can be utilized~\cite{trefethen2020spectra}. First
introduced in the context of black holes~\cite{Jaramillo:2020tuu}, pseudospectrum is usually used to analyze the properties of non-self-adjoint operator in dissipative system, and it provides a visual intuition to the spectral (in)stability of an operator on the complex frequency plane. The behavior of pseudospectrum in terms of BHs in different spacetimes has attracted much attention, the contour lines of $\epsilon-$pseudospectrum will always expand to the infinity based on the overall trend, but there are also some differences. For the asymptotically flat black hole~\cite{Jaramillo:2020tuu,Destounis:2021lum}, the lowest point of a contour line is located at the imaginary axis, when the cosmological constant exists, the location of the lowest point of a contour line depends on the parameters~\cite{Sarkar:2023rhp,Destounis:2023nmb,Arean:2023ejh,Cownden:2023dam,Boyanov:2023qqf}. In addition, the horizonless compact object has its own structure of pseudospectrum~\cite{Boyanov:2022ark} due to the Dirichlet boundary condition. However, this does not imply that the research about a specific perturbation is meaningless, and migrations of QNMs under some specific perturbations are also of great significance~\cite{Courty:2023rxk,Cheung:2021bol,Konoplya:2022pbc}.

Mathematically speaking, equipping with the notion of pseudospectrum, the QNM (in)stability problem reduces into the eigenvalue stability problem associated with a non-selfadjoint operator. For a self-adjoint operator in the non-dissipative system, the spectral theorem guarantees the stability of the spectrum under some perturbations. It means that small perturbations of the operator will also lead to small migrations of eigenvalues. In contrast, for dissipative system such as a black hole, the spectral theorem is not serviceable. This may lead to instabilities. Such kind of instability is manifested by the strong sensitivity of eigenvalues to small perturbations of the operator.

GR proves itself tremendous success with today's precision, but its imperfections such as the singularity problem also motivate physicists to seek for the complete quantum theory of gravity. For Schwarzschild BHs, there are various types of corrections proposed by different theories of gravity.  Loop quantum gravity (LQG) theory is among the candidates that offer such corrections, one of its advantages is to provide an approach to tackle with the singularity issue well. There are some studies on several loop quantum black holes which is coming from LQG. For example, QNMs and shadows are studied on two kinds of loop quantum black holes~\cite{Liu:2021djf,Liu:2020ola}. Recently, the quantum Oppenheimer-Snyder gravitational collapse model has been proposed in the framework of LQG, providing a quantum corrected Schwarzschild spacetime~\cite{Lewandowski:2022zce}. In this work, it is found that the quantum effect means the lower limit of the mass of a black hole generated by a collapsed dust ball. The physical origin of this black hole is different from the solutions which are studied in~\cite{Liu:2021djf,Liu:2020ola}. For this black hole, the spacetime is divided into the interior part and the exterior part. The interior metric is a cosmological metric, and the scaling factor satisfies a deformed Friedmann equation. The interior spacetime is based on the so-called quantum OS model. In other words, it corresponds to the Oppenheimer and Snyder (OS) gravitational collapse model if one does not consider the quantum correction. The exterior metric is static, and its specific form is obtained by certain conditions on boundary surface. The results indicate that the effective exterior spacetime is a quantum corrected Schwarzschild spacetime. Regarding how to connect two parts of the spacetime, one can refer to~\cite{Yang:2022btw}. Another effective framework, using the $\bar{\mu}$ scheme of holonomy corrections and requiring the areal gauge
in the classical theory, results in the same metric~\cite{Kelly:2020uwj}. This quantum corrected black hole exhibits the same asymptotic behavior as the Schwarzschild black hole and it is stable against test scalar and vector fields by the analysis of QNMs~\cite{Yang:2022btw}. Here, ``stable" means that there are no exponential growth modes, which is different from the spectrum stability. The QNMs of other several quantum corrected black holes have been studied in~\cite{Konoplya:2023ahd,Konoplya:2023aph,Konoplya:2022hll,Momennia:2022tug}. Besides the QNMs, other properties of such quantum corrected black hole have also garnered considerable attentions. It has several observable properties that are different from those of the Schwarzschild black hole, including its shadow and photon sphere~\cite{Zhang:2023okw,Ye:2023qks}. The stability of the Cauchy horizon of such black hole  has been discussed in~\cite{Cao:2023aco}. The results support the strong cosmic censorship conjecture (SCCC). By adding the cosmological constant, the authors in~\cite{Shao:2023qlt} focus on checking SCCC. They put forward that SCCC will be destroyed as the black hole approaches the near-extremal limit.

Among the previous works mentioned above, although QNMs have been explored, the pseudospectrum and spectrum (in)stability  are still absent. In this paper, we will study the pseudospectrum and spectrum (in)stability of the quantum corrected Schwarzschild black hole. First, we use the pseudospectrum to derive the spectrum instability which is a general property for black holes. It should be emphasized that the use of pseudospectrum goes far beyond this. Based on the results of pseudospectra, people can further calculate the pseudospectral abscissa and the Kreiss constant. These quantities help us to further understand the transient effect for a time-dependent equation $\mathrm{d}u/\mathrm{d}t=Au$, which may be applied in binary black hole merger~\cite{Jaramillo:2022kuv}. Such transient effect is the one that the QNM spectrum cannot tell us, which is one of the important manifestations of pseudospectrum. We refer the reader to the review~\cite{Destounis:2023ruj} which discusses the pseudospectrum as a newly non-modal tool in BH physics that it can shed light on the spectrum stability of QNMs. 

It is widely believed that the pseudospectrum only provides us the approximate migration direction of the QNM spectrum with a given size of perturbation. However, people may be more interested in how the QNM spectrum migrates with an explicit perturbation form, not just its size. Motivated by this requirement, we then investigate the (in)stability of the spectrum from two aspects as follows. 

On the one hand, we will use this solution as a correction to the Schwarzschild solution to discuss spectrum (in)stability of Schwarzschild black hole, which echoes the previous discussion mentioned about the influence of quantum corrections on the QNM spectrum~\cite{Yang:2022btw,Gong:2023ghh}. This part of investigation will help us to gain insights into the migration behavior of QNMs for the Schwarzschild BH with the aforementioned quantum corrections. By defining the migration ratio, the calculation results manifest that the so-called ``migration ratio instability" will occur between different overtones with regard to the small angular momentum $l$ and large $q$ representing a small black hole.

On the other hand, we regard this black hole as a background spacetime to discuss the spectrum (in)stability of such quantum corrected black hole model who have two horizons like the RN black hole. That is to say, we need to add another specific perturbation into the effective potential to study the impact of different forms of perturbations on spectrum (in)stability. This part of study will equip us to gain insights into the migration behavior of QNMs for such BHs with some external perturbations. It should be noted that perturbation size depends not only on its amplitude, but also on its shape. Fortunately, the concept of pseudospectrum naturally includes characterizations of size~\cite{Jaramillo:2020tuu}. Physically, it is interesting for us to be concerned about the impact of perturbations in terms of the same size but different shapes on the spectrum. Therefore, we study the migrations of spectra along the so-called isoenergetic contour lines, which is different from~\cite{Courty:2023rxk,Cheung:2021bol}. With these motivations in mind, we study the spectrum (in)stability where the perturbation is at the event horizon and null infinity. Under the same energy norm with different quantum corrected parameters, we find several interesting facts. First, perturbations at infinity are more capable of generating spectral instability than those at the event horizon. Second, we find that the peak distribution can lead to instability of QNMs more efficiently than the average distribution. 

The paper is organized as follows. In Sec.\ref{sec: scalar}, we introduce a brief view of the test scalar equation of the quantum corrected black hole. Sec.\ref{sec: set up} offers us some necessary preparations for numerical calculation of pseudospectrum including hyperboloidal framework, energy norm. In Sec.\ref{sec: stability_of_QNMs}, we show the pseudospectrum and discuss the spectrum (in)stability in terms of two different perspectives mentioned above. Sec.\ref{sec: conclusions} is the conclusions and the discussion. In Appendix \ref{app_hyperboloidal_framework}, the hyperboloidal framework is shown. The sophisticated numerical approach is presented in Appendix \ref{app_numerical_approach}. In Appendix \ref{app_Gram_matrix}, we provide the detailed steps of the construction of the Gram matrix $\tilde{\mathbf{G}}^E$. In Appendix \ref{app_subspace_method}, the invariant subspace method is introduced to improve computational efficiency. Methodologically speaking, Appendix \ref{app_Gram_matrix} and Appendix \ref{app_subspace_method} can be seen as supplements to the numerical methods shown in Appendix \ref{app_numerical_approach}. In this paper, our symbol convention is to use bold letters to represent matrices, and un-bold letters to represent functions or operators.

\section{The quantum corrected black hole and the massless scalar field equation}\label{sec: scalar}
In our paper, we  focus on the exterior spacetime of the quantum corrected black hole whose metric is given by~\cite{Lewandowski:2022zce}
\begin{eqnarray}\label{metric}
	\mathrm{d}s^2=-f(r)\mathrm{d}t^2+\frac{\mathrm{d}r^2}{f(r)}+r^2(\mathrm{d}\theta^2+\sin^2\theta\mathrm{d}\phi^2)\, ,
\end{eqnarray}
where the metric function $f(r)$ reads
\begin{eqnarray}\label{metric_function}
	f(r)=1-\frac{2M}{r}+\frac{\alpha M^2}{r^4}\, ,
\end{eqnarray}
with the parameter $\alpha=16\sqrt{3}\pi\gamma^3l^2_p$ , $l_p=\sqrt{\hbar}$ denoting the Planck length, $\gamma$ being the Immirzi parameter, and $M$ standing for the mass of the black hole. By setting $\alpha=0$, we obtain the Schwarzschild spacetime. We use $r=r_{+}$ to denote the event horizon and $r=r_{-}$ to denote the Cauchy horizon. Therefore, the metric function $f(r)$ can be factorized into
\begin{eqnarray}
	f(r)=\Big(1-\frac{r_{+}}{r}\Big)\Big(1-\frac{r_{-}}{r}\Big)\Big(1-\frac{s}{r}+\frac{t}{r^2}\Big)\, ,
\end{eqnarray}
where
\begin{eqnarray}
	s=-\frac{r_{+}r_{-}(r_{+}+r_{-})}{r_{+}^2+r_{+}r_{-}+r_{-}^2}\, ,\quad t=\frac{r_{+}^2r_{-}^2}{r_{+}^2+r_{+}r_{-}+r_{-}^2}\, .
\end{eqnarray}
For a given value of the parameter $\alpha$ and $M$, it can be found that the event horizon $r_{+}$ and the Cauchy horizon $r_{-}$ are uniquely determined by the following relations:
\begin{eqnarray}
	M&=&\frac{r_{+}^3+r_{+}^2r_{-}+r_{+}r_{-}^2+r_{-}^3}{2(r_{+}^2+r_{+}r_{-}+r_{-}^2)}\, ,\\
	\alpha&=&\frac{4r_{+}^3r_{-}^3(r_{+}^2+r_{+}r_{-}+r_{-}^2)}{(r_{+}^3+r_{+}^2r_{-}+r_{+}r_{-}^2+r_{-}^3)^2}\, .
\end{eqnarray}
Moreover, we have a dimensionless expressions $\alpha/M^2$ in terms of $q$ in the following
\begin{eqnarray}\label{alpha_M_q}
	\frac{\alpha}{M^2}=\frac{16q^6(1+q^2+q^4)^3}{(1+q^2+q^4+q^6)^4}\, ,
\end{eqnarray}
where for convenience, we define
\begin{eqnarray}\label{q}
	q^2=\frac{r_{-}}{r_{+}}\ge0\, .
\end{eqnarray}

The Immirzi parameter $\gamma$ can be determined by considering the black hole entropy~\cite{Gan:2022oiy,Meissner:2004ju,Domagala:2004jt}, which results in a restricted range of possible values for parameters $\alpha$ and $q$. It means that $q$ is constrained to be tiny for solar mass black holes, thus the deviation of the solutions studied is negligible in comparing to that of the classical Schwarzschild solution in the viewpoint of LQG. However, there is the same metric form (\ref{metric_function}) coming from different models, which have diverse interpretations and the corresponding corrected parameter can take a larger range. For example, in the framework of the so-called mimetic gravity~\cite{Nashed:2018urj} (see Eq.(51) therein), the metric is interpreted as a natural extension of Schwarzschild metric, and the corresponding corrected parameter is not constrained by the Immirzi parameter $\gamma$. Actually, the coefficient of the quadruple term in the metric (\ref{metric_function}) is an integration constant therein. In other words, the range of parameter $q$ depends on what gravity theory is considered, it is able to take a wider range of values in the mimetic gravity, though it certainly has stringent constraint in LQG. Additionally, it's notable that our subsequent analysis does not incorporate any specific quantum effect in fact and given the interest in theoretical study, it's reasonable to set a larger range of $q$.

Given that this is a quantum effective solution, it is difficult to calculate general perturbations, so we study a test massless scalar field $\Phi$ which satisfies with Klein-Gordon equation as follows
\begin{eqnarray}\label{KG_equation}
	\square\Phi=\frac{1}{\sqrt{-g}}(g^{\mu\nu}\sqrt{-g}\Phi_{,\mu})_{,\nu}=0\, ,
\end{eqnarray}
in which $g$ denotes the determinant of the metric of the background geometry $g_{\mu\nu}$. Considering a standard separation of variables of the form
\begin{eqnarray}
	\Phi(t,r,\theta,\phi)=\sum_{l,m}\frac{\Psi_{l,m}(t,r)}{r}Y_{l,m}(\theta,\phi)\, ,
\end{eqnarray}
with $Y_{l,m}$ standing for the spherical harmonic function, Eq.(\ref{KG_equation}) is rewritten as 
\begin{eqnarray}\label{wave_equation}
	\Big(\frac{\partial^2}{\partial t^2}-\frac{\partial^2}{\partial r_{\star}^2}+V_l\Big)\Psi_{l,m}=0\, ,
\end{eqnarray}
where the tortoise coordinate $r_{\star}$ and the effective potential $V_l$ are defined by 
\begin{eqnarray}\label{tortoise_coordinate_and_effective_potential}
	\mathrm{d}r_{*}=\frac{\mathrm{d}r}{f(r)}\, ,\quad V_l(r)=f(r)\Big[\frac{l(l+1)}{r^2}+\frac{f^{\prime}(r)}{r}\Big]\, ,
\end{eqnarray}
and $l=0,1,2,3,\dots$ is referred to a angular momentum number. For convenience, we omit $l$ and $m$ for $\Psi_{l,m}$ in following paper.

Our interesting region is a patch between the event horizon $r_{+}$ and the spatial infinity $r\to\infty$. QNMs are the solutions of the Eq.(\ref{wave_equation}) that satisfy a purely ingoing boundary condition at the event horizon, i.e.,
\begin{eqnarray}\label{event_boundary_condition}
	(\Psi_{,t}-\Psi_{,r_{\star}})|_{r=r_{+}}=0\, ,
\end{eqnarray}
and a purely outing boundary condition at the infinity, i.e.,
\begin{eqnarray}\label{infinity_boundary_condition}
	(\Psi_{,t}+\Psi_{,r_{\star}})|_{r\to \infty}=0\, .
\end{eqnarray}
After imposing a Fourier decomposition $\Psi(t,r_{\star})\sim\Psi(r_{\star})e^{i\omega t}$ and the boundary conditions (\ref{event_boundary_condition})-(\ref{infinity_boundary_condition}), we will get an eigenvalue problem. From our conventions, the solution decays over time only when QNM frequencies $\omega_n$ are supposed to have a positive imaginary part. Nevertheless, in the Schwarzschild-like coordinate, there is a divergent behavior in the spatial part of the QNMs. It is precise for this reason that when we perform numerical methods for calculating the QNM frequencies, we need to additionally extract the divergent behavior of the solution. These numerical methods include, but are not limited to, the continued fractions method~\cite{Leaver:1985ax}, the asymptotic iteration method~\cite{Cho:2011sf}, the pseudo-spectral method with the Chebyshev polynomial~\cite{Mamani:2022akq}, the pseudo-spectral method with the Bernstein polynomial~\cite{Fortuna:2020obg} and the matrix method~\cite{Lin:2016sch}. We will resolve this issue by switching to the hyperboloidal slicing, which is described in~\cite{Jaramillo:2020tuu} in the study of the QNMs.

\section{The set up}\label{sec: set up}
\subsection{The first-order reduction}
Following the Ref.\cite{Jaramillo:2020tuu} with the techniques described in the Ref.\cite{PanossoMacedo:2023qzp}, we present a hyperboloidal coordinate system that imposes the QNMs boundary conditions geometrically. The hyperboloidal coordinate $(\tau,\sigma,\theta,\phi)$ is
\begin{eqnarray}\label{hyperboloidal_coordinate}
	t=\lambda\Big(\tau-H(\sigma)\Big)\, ,\quad r=\lambda\frac{\rho(\sigma)}{\sigma}\, .
\end{eqnarray}
Here, $\lambda$ is a given characteristic length scale of the spacetime, and the function $H(\sigma)$ is called the height function, which is constructed by requiring that $\tau=\text{constant}$ slices are spacelike hypersurfaces penetrating the future event horizon and the future null infinity. The construction of a general family of hyperboloidal coordinates suitable to describe black hole perturbations has been put forward in~\cite{Zenginoglu:2007jw}. The paper further explains the relationship among time transformation, spatial compactification, and conformal rescaling that is crucial for the construction of the minimal gauge~\cite{PanossoMacedo:2023qzp}, which is a specific member of the family of hyperboloidal coordinates. For simplicity, the minimal gauge is performed to find the height function in our paper. However, as the Ref.\cite{PanossoMacedo:2023qzp} stated, there are two approachs to determine the height function. One is called the in-out strategy and the other is called the out-in strategy. After our analysis, it is found that both strategies are viable for constructing hyperboloidal coordinates. For convenience, we adopt the in-out strategy. Hereafter, we will omit the markings ``in-out" unless clarification is needed. One can find more details in Appendix \ref{app_hyperboloidal_framework}. 

Now, in order to calculate the spectrum and pseudospectrum, we follow the Ref.\cite{Jaramillo:2020tuu} and cast the QNM eigenvalue problem in the hyperboloidal scheme. After a reduction of order in time $\Pi=\partial_\tau\Psi$, Eq.(\ref{wave_equation}) can be rewritten as two partial differential equations, involving first-order derivative respect to time and second-order derivative respect to space., i.e.,
\begin{eqnarray}\label{pde}
	\partial_\tau u=iLu\, ,\quad L=\frac{1}{i}\begin{bmatrix}
		0 & 1\\
		L_1 & L_2
	\end{bmatrix}\, ,\quad u=\begin{bmatrix}
		\Psi\\
		\Pi
	\end{bmatrix}\, ,
\end{eqnarray}
where two components $L_1$ and $L_2$ of the operator $L$ are given by
\begin{eqnarray}\label{operator_L_1_L_2}
	L_1&=&\frac{1}{w(\sigma)}\Big[\partial_\sigma(p(\sigma)\partial_{\sigma})-q_{l}(\sigma)\Big]\, ,\nonumber\\
	L_2&=&\frac{1}{w(\sigma)}\Big[2\gamma(\sigma)\partial_\sigma+\partial_\sigma\gamma(\sigma)\Big]\, .
\end{eqnarray}
Here, the four functions $w(\sigma)$, $p(\sigma)$, $q_{l}(\sigma)$ and $\gamma(\sigma)$ are written as 
\end{multicols}
\noindent\rule{85.5mm}{0.4pt}\rule{0.4pt}{2mm}
\begin{eqnarray}\label{function_w_p_q_gamma}
w(\sigma)=\frac{1-\gamma(\sigma)^2}{p(\sigma)}\, ,\quad p(\sigma)=\sigma^2f(\sigma)\, ,\quad q_{l}(\sigma)=\frac{\lambda^2}{p(\sigma)}V_l(\sigma)\, ,\quad \gamma(\sigma)=H^{\prime}(\sigma)p(\sigma)\, ,
\end{eqnarray}
\noindent\hspace{92.5mm}\vrule width0.4pt height0.4pt depth 2mm\rule{85.5mm}{0.4pt}
\begin{multicols}{2}
where the metric function $f$ and the potential function $V_l$ have been built in terms of the coordinate $\sigma$. As a result, the four functions containing $q$ as a parameter appearing at Eqs.(\ref{operator_L_1_L_2}) turn into the forms as follows
\end{multicols}
\noindent\rule{85.5mm}{0.4pt}\rule{0.4pt}{2mm}
\begin{eqnarray}\label{function_w}
w(\sigma)&=&-\frac{4}{\left(q^4+q^2+1\right)^3}\Big[q^{18}\sigma(\sigma ^3-1)+q^{16}(2 \sigma ^4+2\sigma^3-3\sigma-1)+q^{14}(3 \sigma ^4+4 \sigma ^3+\sigma ^2-6 \sigma -3)\nonumber\\
&&+2 q^{12}(2 \sigma ^4+3 \sigma ^3+\sigma ^2-5 \sigma -3)+3 q^{10}(\sigma ^4+2 \sigma ^3+\sigma ^2-4 \sigma -3)+2 q^8 (\sigma ^4+2 \sigma ^3+\sigma ^2-6 \sigma -5)\nonumber\\
&&+q^6 (\sigma ^4+2 \sigma ^3+\sigma ^2-10 \sigma -9)-6 q^4 (\sigma +1)-3 q^2 (\sigma +1)-\sigma -1\Big]\, ,
\end{eqnarray}
\begin{eqnarray}\label{function_p}
p(\sigma)=\frac{(\sigma -1) \sigma ^2(q^2 \sigma -1)\Big[q^4 (\sigma ^2+\sigma +1)+q^2 (\sigma +1)+1\Big]}{q^4+q^2+1}\, ,
\end{eqnarray}
\begin{eqnarray}\label{function_q}
q_l(\sigma)=l^2+l+\frac{\sigma \Big[q^6 (1-4 \sigma ^3)+q^4+q^2+1\Big]}{q^4+q^2+1}\, ,
\end{eqnarray}
\begin{eqnarray}\label{function_gamma}
\gamma(\sigma)&=&\frac{1}{(q^4+q^2+1)^2}\Big[2 q^{12} \sigma ^2 (\sigma ^3-1)+2 q^{10} \sigma ^2 (\sigma ^3+\sigma ^2-2)+q^8 (2 \sigma ^5+2 \sigma ^4-6 \sigma ^2+1)\nonumber\\
&&+2 q^6(\sigma ^5+\sigma ^4-4 \sigma ^2+1)+q^4 (3-6 \sigma ^2)+q^2(2-4 \sigma ^2)-2 \sigma ^2+1\Big]\, .
\end{eqnarray}
\noindent\hspace{92.5mm}\vrule width0.4pt height0.4pt depth 2mm\rule{85.5mm}{0.4pt}
\begin{multicols}{2}
Performing the Fourier transformation in $\tau$  
\begin{eqnarray}
	u(\tau,\sigma)\sim u(\sigma)e^{i\omega\tau}\, ,
\end{eqnarray}
we arrive at the eigenvalue problem as follows
\begin{eqnarray}\label{QNM_eigenvalue_problem}
	\frac{1}{i}\begin{bmatrix}
		0 & 1\\
		L_1 & L_2
	\end{bmatrix}
	\begin{bmatrix}
		\Psi\\
		\Pi
	\end{bmatrix}=\omega
	\begin{bmatrix}
		\Psi\\
		\Pi
	\end{bmatrix}\, .
\end{eqnarray}
Note that for any $\tau$ defined in Eq.(\ref{hyperboloidal_coordinate}), the corresponding frequencies $\omega$ conjugating to $t$ and $\tau$ coincide up to the rescaling constant $\lambda$. The rescaling constant  is equal to the event horizon $r_{+}$ in our set up. Specifically, we have $\omega_{\tau}=\lambda\omega_{t}$, in which $\omega_{\tau}$ is derived in the $\tau$-picture and $\omega_t$ is derived in the $t$-picture. This formula will be used to check the reliability of our calculations for QNM frequencies.

\subsection{The pseudospectrum and the energy norm}
Up to now, we have obtained the QNM eigenvalue problem (\ref{QNM_eigenvalue_problem}) in the hyperboloidal frame for the sake of calculating the spectrum and pseudospectrum. 
Given $\epsilon>0$, the $\epsilon$-pseudospectrum $\sigma_{\epsilon}(A)$ of an operator $A$ is defined as~\cite{trefethen2020spectra}
\begin{eqnarray}\label{pseudospectrum_definition}
	\sigma_{\epsilon}(A)&=&\{z\in\mathbb{C}:\lVert R_{A}(z)\rVert=\lVert(z\mathbb{I}-A)^{-1}\rVert>1/\epsilon\}\, ,
\end{eqnarray}
where $R_{A}(z)$ is called the resolvent operator. This definition is the most suitable for visualizing the pseudospectrum. In the limit $\epsilon\to0$, the set $\sigma_\epsilon(A)$ reduces to the spectrum set $\sigma(A)$, whose elements are the spectrum $\omega_n$, in which $n$ is called the overtone number. The quantity $\epsilon$ serves as a measure of the ``proximity" between points in $\sigma_\epsilon(A)$ and the spectrum $\omega_n$, offering a clear interpretation of perturbations to the underlying operator. Therefore, the shape and size of the $\epsilon$-pseudospectrum regions  quantify the spectrum (in)stability of the operator $A$. In practice, if the contour lines of the pseudospectrum forms a concentric circle of the spectrum, then the operator considered is spectrally stable. On the contrary, if this structure does not occur, then this operator is spectrally unstable. It is known that with regard to the QNMs, the spectrum is generally unstable since the black hole is actually a leaky system~\cite{Jaramillo:2020tuu,Destounis:2021lum,Arean:2023ejh,Cownden:2023dam,Boyanov:2023qqf,Sarkar:2023rhp,Destounis:2023nmb}. However, some exceptions exist. For instance, consider the massless scalar perturbations in global $\text{AdS}_4$ spacetime where the energy is conservative, and therefore the spectrum is stable~\cite{Cownden:2023dam}.

\begin{figure}[H]
	\centering
	\includegraphics[scale=0.7]{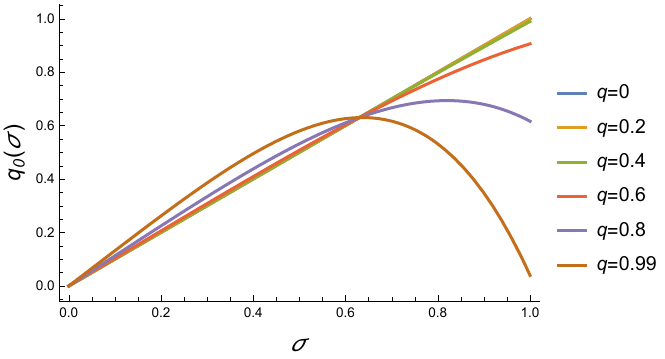}
	\caption{The functions $q_l(\sigma)$ for the quantum corrected black hole with different $q$ and $l=0$ in the interval $[0,1]$, in which $q_l(\sigma)$ is defined in Eq.(\ref{function_q}) and $q$ is defined in Eq.(\ref{q}).}
	\label{fig: ql}
\end{figure}

From the definition (\ref{pseudospectrum_definition}) of pseudospectrum, it can be found that the pseudospectrum depends on the norm chosen but the spectrum does not. Norm is a mathematical concept that can have many different choices. Its form will also affect what is meant by ``small" or ``large" perturbations. Physically, we use the energy norm as the norm for calculation~\cite{Jaramillo:2020tuu,Gasperin:2021kfv}. The definition of this norm associated with the field $\Psi$ and $\Pi$ in the asymptotically flat spacetime is given by
\end{multicols}
\noindent\rule{85.5mm}{0.4pt}\rule{0.4pt}{2mm}
\begin{eqnarray}\label{energy_norm_sigma}
\lVert u\rVert_{E}^2&=&\Big\lVert \begin{bmatrix}
	\Psi\\
	\Pi
\end{bmatrix}\Big\rVert_E^2=\frac12\int_{0}^1\Big[w(\sigma)|\Pi|^2+p(\sigma)|\partial_\sigma\Psi|^2+q_l(\sigma)|\Psi|^2\Big]\mathrm{d}\sigma\nonumber\\
&=&\frac12\int_{0}^1\Big[w(\sigma)\Pi^{\star}\Pi+p(\sigma)\partial_\sigma\Psi^{\star}\partial_\sigma\Psi+q_l(\sigma)\Psi^{\star}\Psi\Big]\mathrm{d}\sigma\, .
\end{eqnarray}
\noindent\hspace{92.5mm}\vrule width0.4pt height0.4pt depth 2mm\rule{85.5mm}{0.4pt}
\begin{multicols}{2}
Here, the symbol $\star$ stands for the complex conjugate and functions $w(\sigma)$, $p(\sigma)$, $q_l(\sigma)$ are given by Eqs.(\ref{function_w})-(\ref{function_q}), respectively. The definition of this norm is also suitable for our model. It should be noted that in order to keep the positive definiteness of the energy norm in the interval $[0,1]$, all three functions $w(\sigma)$, $p(\sigma)$, $q_l(\sigma)$ should be positive definite\footnote{The positive definiteness is quite important. Recently, it is found that in the dS case with $l=0$, the Gram matrix is singular~\cite{Destounis:2023nmb}. Therefore, the adjoint operator $L^{\dagger}$ can be ill-defined.}. The positive definiteness of the function $w(\sigma)$, i.e., 
$[\gamma(\sigma)]^2<1$ has been illustrated in Appendix \ref{app_hyperboloidal_framework}. The positive definiteness of the function $p(\sigma)$ is obvious. Furthermore, we show the positive definiteness of the functions $q_l(\sigma)$. For $l=0$, we see that $q_0(\sigma)\ge0$ in the Fig.\ref{fig: ql}. We can observe that $q_l(\sigma)$ satisfy $0\le q_0(\sigma)\le q_1(\sigma)\le q_2(\sigma)\le q_3(\sigma)\le \cdots$. Therefore, for all $l\ge0$, we have $q_l(\sigma)\ge0$.


With the definition of energy norm (\ref{energy_norm_sigma}), the definition of the inner product is produced by
\end{multicols}
\noindent\rule{85.5mm}{0.4pt}\rule{0.4pt}{2mm}
\begin{eqnarray}\label{energy_inner_product_sigma}
\langle u_1,u_2\rangle_E=\Big\langle\begin{bmatrix}
	\Psi_1\\
	\Pi_1
\end{bmatrix}\, ,
\begin{bmatrix}
	\Psi_2\\
	\Pi_2
\end{bmatrix}\Big\rangle_E=\frac12\int_{0}^1\Big[w(\sigma)\Pi_1^{\star}\Pi_2+p(\sigma)\partial_\sigma\Psi_1^{\star}\partial_\sigma\Psi_2+q_l(\sigma)\Psi_1^{\star}\Psi_2\Big]\mathrm{d}\sigma\, .
\end{eqnarray}
\noindent\hspace{92.5mm}\vrule width0.4pt height0.4pt depth 2mm\rule{85.5mm}{0.4pt}
\begin{multicols}{2}
When $u_1=u_2$, Eq.(\ref{energy_inner_product_sigma}) reduces to Eq.(\ref{energy_norm_sigma}). In the next subsection, we provide a detailed introduction to the numerical technique of using the energy norm defined above to determine the pseudospectrum, which in turn will inform us of the stability of the QNM frequencies for the quantum corrected black hole.

\section{(In)stability of the QNM spectrum}\label{sec: stability_of_QNMs}
Now, the tools needed for computation have been established, then we will discuss the stability of QNM spectrum for the quantum corrected black hole in this section. The $\epsilon$-pseudospectrum of the black hole can be interpreted as a topographic map that characterizes the stability of QNM spectrum. As we have emphasized before, if the spectrum is stable under external disturbances, then the pseudospectrum will have a concentric structure. The contour lines around a specific eigenvalue will resemble circles with a radius of $\epsilon$, which corresponds to the intensity of the disturbance. Conversely, when away from the spectrum, any non-trivial topographic structure extending towards the complex plane will be a sign of instability. The predictive ability of $\epsilon$-pseudospectrum lies in that it can estimate the potential changes for the QNM spectrum under the influence of the scattering potential $V$ which experiences a perturbation $\delta V$. Roughly speaking, pseudospectrum can provide perception on how much a given size of perturbation can influence the spectrum of an operator.

\begin{figure*}
	\centering
	\includegraphics[scale=0.4]{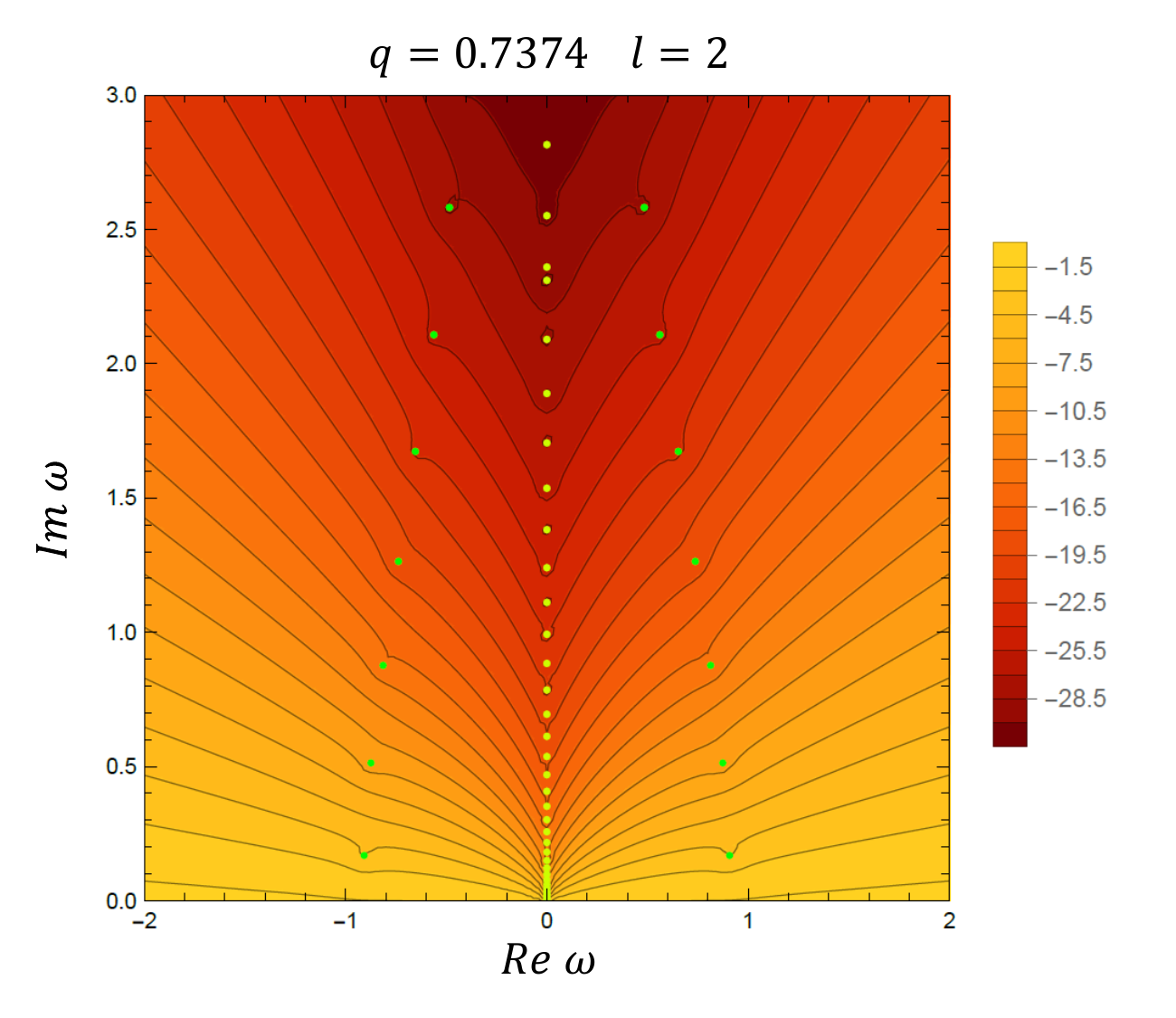}
	\includegraphics[scale=0.4]{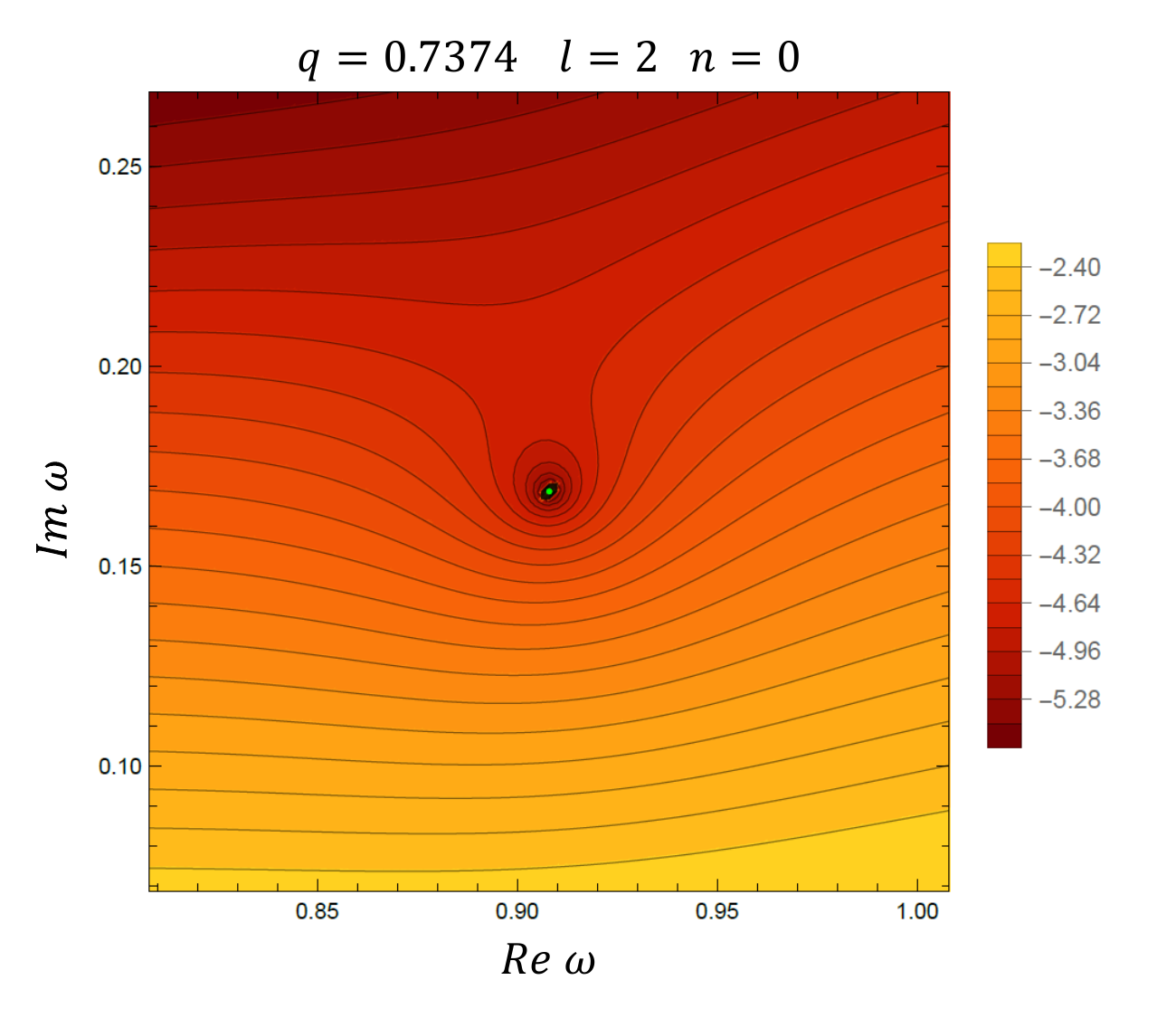}\\
	\includegraphics[scale=0.4]{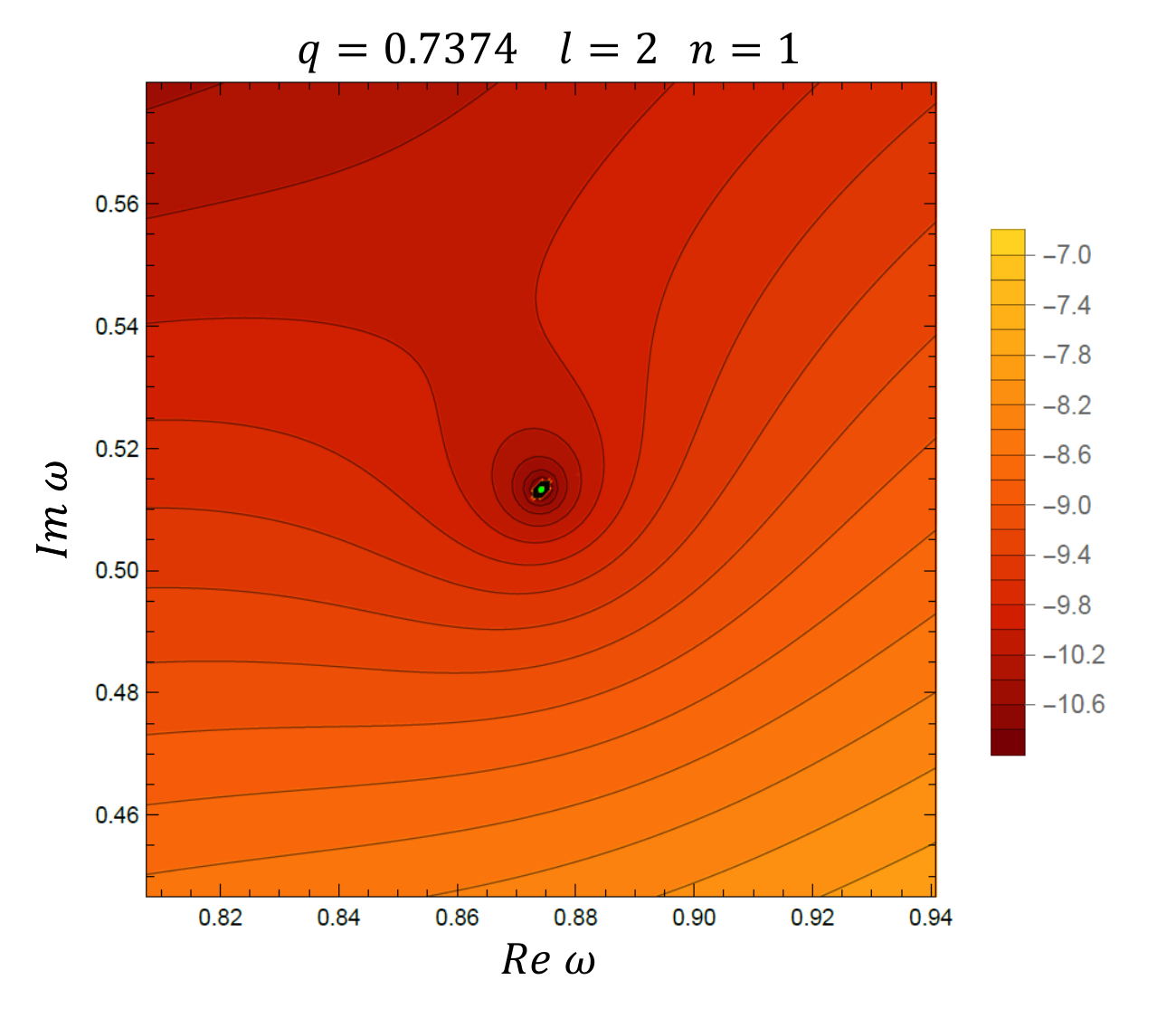}
	\includegraphics[scale=0.4]{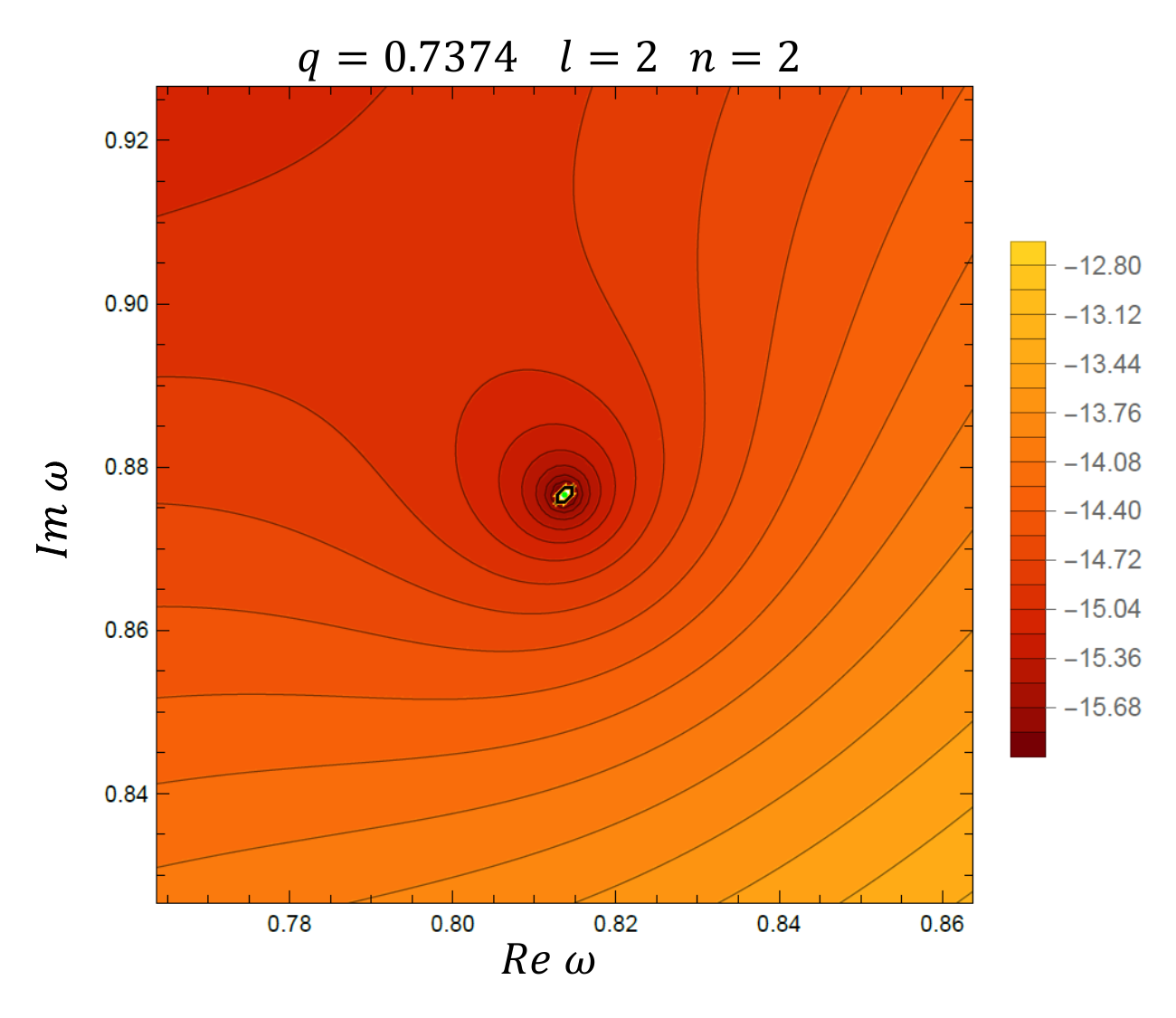}
	\caption{The $\epsilon$-pseudospectrum of the quantum corrected black hole with $q=0.7374$ and $l=2$. The solid contour lines correspond to various choices of $\log_{10}\epsilon$. The calculations have been implemented by using $N=150$ and $I_{\text{max}}=18$. The top left panel represents the zoomed-out view of the $\epsilon$-pseudospectrum. The top right panel represents the zoomed-in view of the $\epsilon$-pseudospectrum around the fundamental mode $n=0$. The bottom left panel represents the zoomed-in view of the $\epsilon$-pseudospectrum around the first overtone $n=1$. The bottom right panel represents the zoomed-in view of the $\epsilon$-pseudospectrum around the second overtone $n=2$.}
	\label{fig: pseudospectrum_example}
\end{figure*}

As for our case belonging to the asymptotic flatness type, the pseudospectrum structure is similar to previous works. As a typical example, we show the pseudospectrum and the spectrum of the quantum corrected black hole with $q=0.7374$ in Fig.\ref{fig: pseudospectrum_example}. The green points are the quasinormal modes and the yellow points are the branch cut modes. These branch cut modes do not converge, and as the number of grid points $N$ increases, these points will be densely distributed on the imaginary axis. The selection of this parameter corresponds exactly to the parameter selection in~\cite{Yang:2022btw}. In this paper, one can get the event horizon $r_{+}=1.8393$. By the hyperboloidal framework, we find the fundamental characteristic frequency is $\omega_{\tau0}=0.9080+0.1687i$. Then in the $t$-picture mentioned in the previous section, we have $\omega_{t0}=\omega_{\tau0}/r_{+}=0.4937+0.09175i$. This result is consistent with~\cite{Yang:2022btw}. This also demonstrates the correctness of using hyperboloidal frameworks to solve QNM problems from a methodological perspective.

From these figures, it can be found that when we arbitrarily zoom in the spectrum, a circular structure will still appear. However, the contour lines still have an open structure from a large-scale perspective. This is precisely the characteristic of the pseudospectrum of a non-self adjoint operator. In fact, we cannot determine if the corresponding operator is self adjoint based solely on whether there is a closed circular structure in the contour lines of the pseudospectrum. As far as we know, to determine whether the operator is self-adjoint or non-selfadjoint, it is important to observe the intersection of closed circular lines surrounding adjacent spectrum points as $\epsilon$ gradually increases, if the integrity of this circular structure is preserved, the operator is self-adjoint, in contrast, if the integrity is damaged, the operator is non-selfadjoint.

Furthermore, for those contour lines that form enclosed circles around the QNMs, we observe that the larger overtone number $n$ is, the smaller the radius of the circle is (see Fig.\ref{fig: pseudospectrum_example}). Roughly speaking, when the intensity $\epsilon$ of perturbation for the operator is greater than the radius of the outermost circle, the corresponding spectrum will migrate into regions that are much further away than $\epsilon$. In other words, the size of the outermost circle serves as a qualitative indication of the stability of corresponding spectrum.

There are two ways to consider the spectrum (in)stability in terms of our model. The first way is to consider the quantum corrected solution as a correction to the Schwarzschild solution, and the corresponding effective potential will also have some changes. The difference between our newly obtained operator $L(x,q)$ and the Schwarzschild solution operator $L(x,0)$ is caused by $q$, i.e., the parameter $\alpha$ [see the relation (\ref{alpha_M_q}) between $\alpha$, $M$ and $q$], leading to the migration of the new spectra relative to the Schwarzschild QNM spectra. In other words, we consider whether the correction of the Schwarzschild solution caused by the quantum correction will trigger large-scale spectrum migration or not. To some extent, the current considerations are the same as directly calculating the QNMs of such quantum corrected solution~\cite{Yang:2022btw,Gong:2023ghh}. The first subsection of present section introduces some results and discussions.

The second way considered is to add a perturbation $\delta q_l$ into the effective potential on the background of the quantum corrected black hole. At present, this perturbation approach, which is used to study the stability of the QNM spectrum, is considered among a number of works~\cite{Jaramillo:2020tuu,Sarkar:2023rhp,Arean:2023ejh,Cownden:2023dam,Boyanov:2023qqf,Destounis:2023nmb}. In doing so, the perturbation of the operator is entirely caused by the perturbation of the effective potential [see Eq.(\ref{delta_L_perturbation_V}) below]. Furthermore, this approach implies that the perturbation of the effective potential does not alter the geometry of spacetime, which is different from the first case. In the second subsection of this section, we consider perturbations at the event horizon and null infinity, respectively. Importantly, sinusoidal perturbations are no longer considered, and the situation for equal energies of the perturbation is being examined. Along the isoenergetic parametric curve, we study the migration and (in)stability of spectra. Detailed calculations and discussions are presented in the second subsection of this section.

\subsection{Perturbations caused by the quantum correction}\label{sub:A}
In this subsection, we concentrate on the modified operator $L(x,q)$, which is caused by the quantum corrected parameter $\alpha$ for the operator $L(x,0)$, and calculate the QNMs corresponding to the modification. One can find the concrete form of $L(x,q)$ in Eq.(\ref{L1_x_q}) and Eq.(\ref{L2_x_q}). There is something important to clarify. Perhaps, someone naively thinks that the so-called ``modified operator" $\delta L$ caused by the quantum correction is written as
\begin{eqnarray}\label{delta_L_quantum_correction}
	\delta L(x,q)&=&L(x,q)-L(x,0)\, .
\end{eqnarray}
But what we need to point out is that, the ``modified operator" $\delta L$ is not a well-defined operator. The reasons are given as follows. The eigenvalue problems corresponding to two operators are expressed as
\begin{eqnarray}
	&&L(x,q)\begin{bmatrix}
		\Psi^q_n\\
		\Pi^q_n
	\end{bmatrix}=\omega^q_n
	\begin{bmatrix}
		\Psi^q_n\\
		\Pi^q_n
	\end{bmatrix}\, ,\quad\text{and}\nonumber\\
	&&L(x,0)\begin{bmatrix}
		\Psi^0_n\\
		\Pi^0_n
	\end{bmatrix}=\omega^0_n
	\begin{bmatrix}
		\Psi^0_n\\
		\Pi^0_n
	\end{bmatrix}\, ,
\end{eqnarray}
respectively, and $n$ is used to mark eigenvalues. However, the domains of the eigenvector functions $[\Psi^q_n,\Pi^q_n]^T$ and $[\Psi^0_n,\Pi^0_n]^T$ are different, i.e., two event horizons $r^q_{+}$ and $r^0_{+}$ are different. In other words, the hyperboloidal coordinate transformation (\ref{hyperboloidal_coordinate}) depend on $q$.  Therefore, Eq.(\ref{delta_L_quantum_correction}) is meaningless. One possible solution that avoids the hyperboloidal framework method used in this paper is directly considering the original version of the eigenvalue problem. Such eigenvalue problem can be written as
\begin{eqnarray}
	\Big(-\frac{\mathrm{d}^2}{\mathrm{d}r_{\star}^2}+V_l(r_{\star},q)\Big)\Psi(r_{\star})=\omega^2\Psi(r_{\star})\, ,
\end{eqnarray}
with the ingoing and outgoing boundary conditions. The property of perturbation to such operator is discussed in the monograph~\cite{sjostrand2019non}. This goes beyond the scope of our present work. 

Anyway, one can always study the migration of the QNM spectrum without introducing the perturbation operator. More precisely, following Refs.\cite{Cheung:2021bol,Courty:2023rxk,Destounis:2023nmb}, we are going to use the migration ratio defined as
\begin{eqnarray}\label{delta_omega}
	\delta \omega^{q}_n=\frac{|\omega_n^{q}-\omega_n^{0}|}{|\omega_n^{0}|}\, ,
\end{eqnarray}
to quantify the relative migration distance between the unperturbed mode and the perturbed mode, where $n$ stands for the overtone number and $0$ refers to the unperturbed QNMs derived in the Schwarzschild black hole here. It should be noted that the sort of overtone is still based on the absolute value of the imaginary part. In this definition, the fundamental mode is considered as the mode with the smallest imaginary part. Here, the migration of the QNMs are originated from the quantum correction $\alpha$, i.e., the dimensionless parameter $q$.

\begin{figure*}
	\centering
	\includegraphics[scale=0.5]{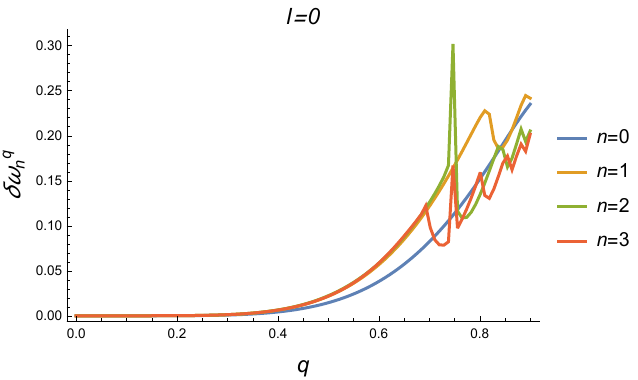}
	\hspace{0.5cm}
	\includegraphics[scale=0.5]{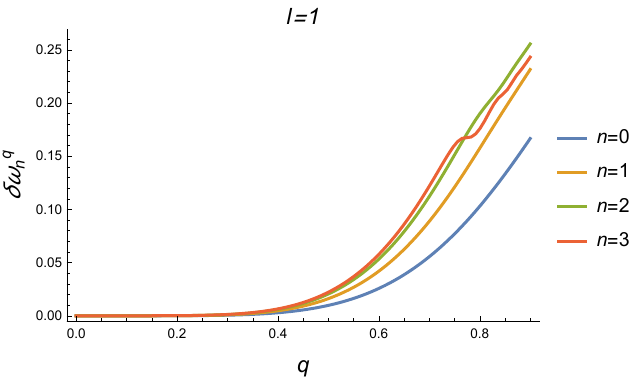}
	\hspace{0.5cm}
	\includegraphics[scale=0.5]{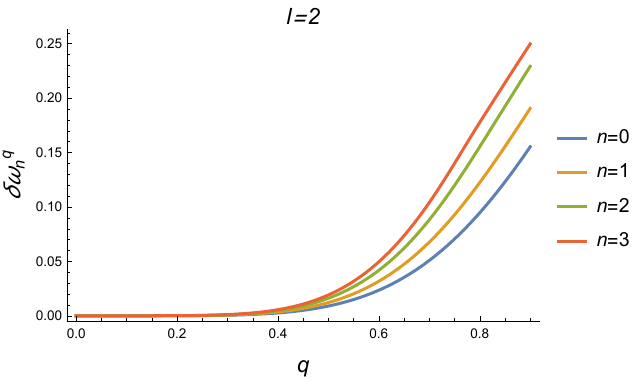}
	\includegraphics[scale=0.5]{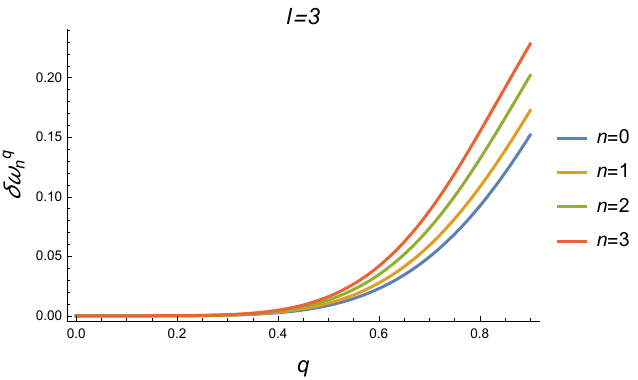}
	\hspace{0.5cm}
	\includegraphics[scale=0.5]{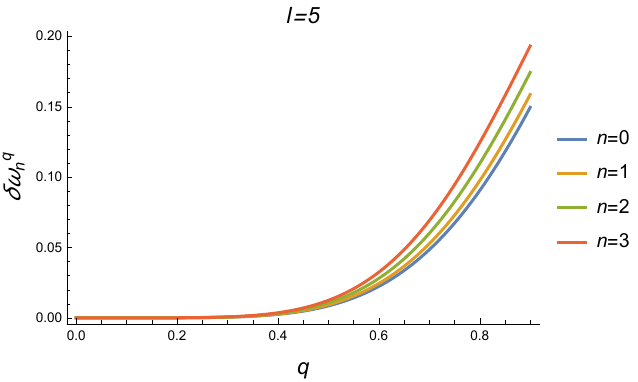}
	\hspace{0.5cm}
	\includegraphics[scale=0.5]{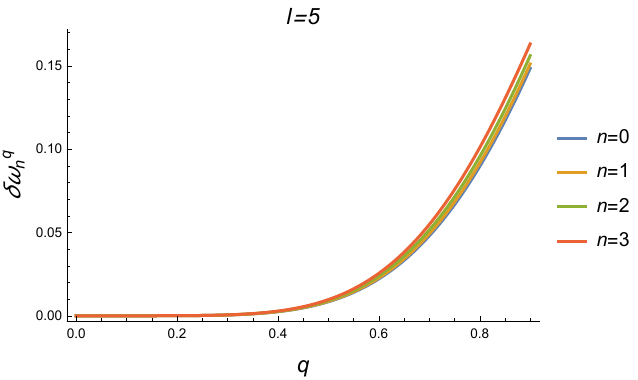}
	\includegraphics[scale=0.5]{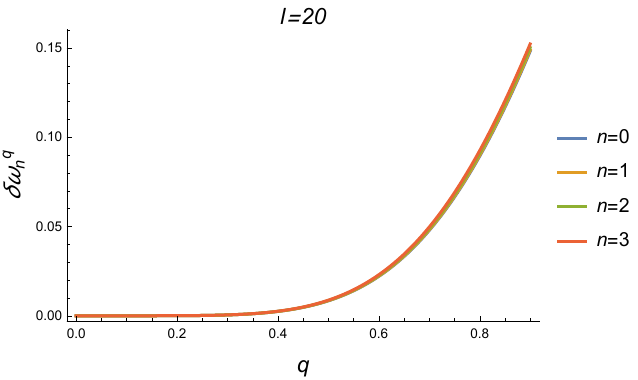}
	\hspace{0.5cm}
	\includegraphics[scale=0.5]{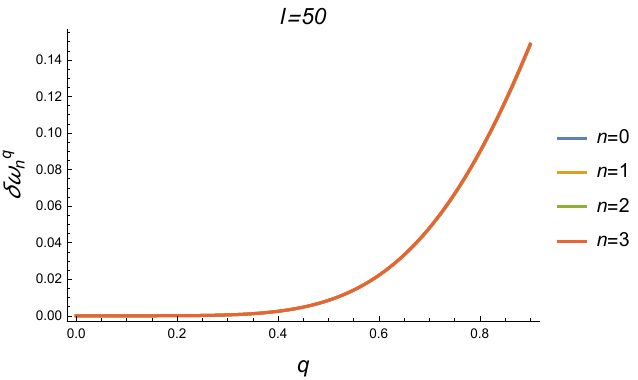}
	\hspace{0.5cm}
	\includegraphics[scale=0.5]{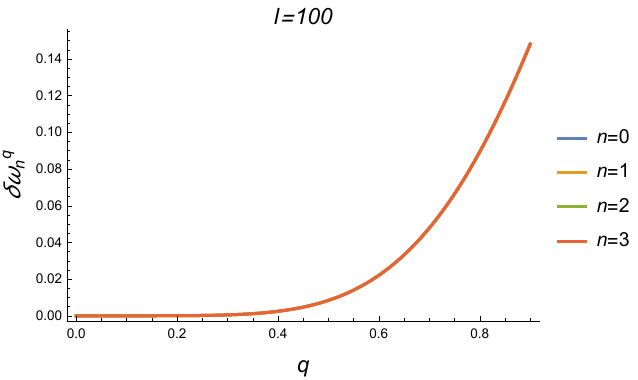}
	\caption{The migration ratio $\delta \omega^{q}_n$ for $l=0$, $1$, $2$, $3$, $5$, $10$, $20$, $50$ and $100$ with different overtones $n=0$, $1$, $2$ and $3$. The number of grid points used is $N=150$. The computation spacing for the parameter $q$ is $\Delta q=0.009$.}
	\label{fig: migration_q}
\end{figure*}

Now, we will describe how the migration ratio is affected by $q$ with respect to different $l$ and $n$. Considering that the small nature of quantum correction in reality, the parameter $q$ should be rather small. But for theoretical research, a larger range of $q$ is more interesting for us. Therefore, we choose the parameter $q$ ranged on the interval $[0,0.9]$. The case of large $q$ corresponds to the case of small black holes. For different angular quantum numbers $l$, the results for the migration ratio are shown in Fig.\ref{fig: migration_q}. From these figures, we summarize the following observations. First, as $q$ increases, the overall migration ratios increase. Except for the cases $l=0$ and $l=1$, the migration ratio increases as $n$ increases with regard to the same $l$, which aligns with our stereotype of high overtone modes. Second, it is found that the angular momentum number has ability to reduce the migration ratio. It means that as $l$ increases, the migration ratio decreases. Interestingly, in the eikonal limit $l\to\infty$, the migration ratio does not change with the change of overtone number $n$. This is natural, because in the eikonal limit $l\to\infty$, one has the characteristic frequencies in the $\tau$-picture~\cite{Cardoso:2008bp}
\begin{eqnarray}
	\omega^q_n=r_{+}^q\Big[\Omega_cl+i\Big(n+\frac12\Big)|\lambda_{L}|\Big]\approx r_{+}^q\Omega_cl\, ,
\end{eqnarray}
which does not depend on the overtone number $n$. Here, $\Omega_c$ is the angular velocity and $\lambda_L$ is the Lyapunov exponent. Last but not least, we see some oscillatory behaviors of migration ratios for the cases $l=0$ and $l=1$. Especially for the case $l=0$, the oscillatory behavior is particularly intense. For these four modes $n=0$, $n=1$, $n=2$ and $n=3$, the ranking of migration ratios will alternate as $q$ increases. The phenomenon begins to occur when $q\approx0.7$ or $q\approx0.77$ for $l=0$ or $l=1$. The similar oscillatory behaviors have also been found in~\cite{Gong:2023ghh}, and the values of $q$ corresponding to the peaks in case $l=0$ are also consistent with the work. Furthermore, we propose that such an oscillatory behavior is the ``migration ratio instability", which occurs between different overtones as $q$ approaches to $1$ that represents the black hole becoming smaller. The reason is stated in~\cite{Konoplya:2022pbc}, in which they use a general method to show that the deformations of the near-horizon parameters such the quantum corrections will induce outburst of first few overtones except for the fundamental mode.

\subsection{Perturbations in the potential $V_l$}\label{sub:B}
Although pseudospectrum allows us to visualize the (in)stability of QNM spectrum under generic perturbations\footnote{Recently, the authors in~\cite{Boyanov:2023qqf} study the convergence issues of the pseudospectrum. They find that although the contour lines at a given resolution $N$ capture the
	qualitative spectrum instability, the non-convergence with $N$ of their associated $\epsilon$ values prevents any quantitative conclusion from the pseudospectrum.}, we are also fascinated by how specific perturbations affect the spectrum~\cite{Cheung:2021bol,Courty:2023rxk}. These perturbations typically occur in the effective potential 
$V_l$ and can also take any form in general, but more importantly, they are disturbances generated by some realistic physical processes. In fact, the most universal perturbation is for the operator $L$, not only for the effective potential. In terms of the previous subsection, new operator $L(x,q)$ is caused by the deviations from pure Schwarzschild geometry due to the quantum correction parameter $\alpha$.

\begin{figure*}
	\centering
	\includegraphics[scale=0.7]{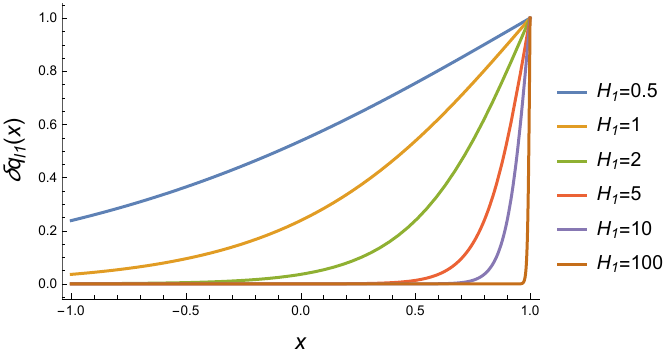}
	\hspace{1cm}
	\includegraphics[scale=0.7]{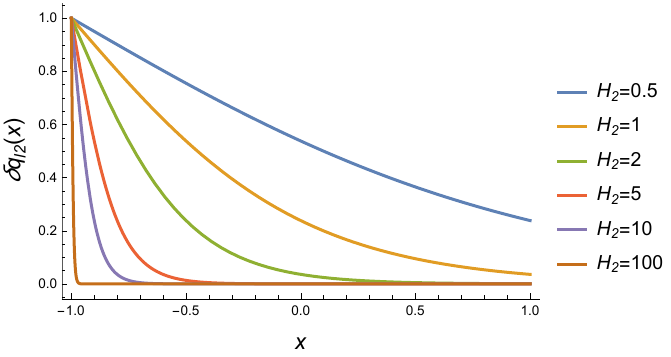}
	\caption{The left panel stands for the functions $\delta q_{l1}(x)$ and the right panel stands for the function $\delta q_{l2}(x)$.  The amplitude in both case is satisfied with $A_1=A_2=1$. }
	\label{fig: deltaql}
\end{figure*}

Now, in this subsection, we focus on a particular case. Namely, the perturbation of operator $\delta L$ is only generated by the perturbation of the effective potential $\delta q_l$ (Here, in $x$ coordinate, we use $\delta q_l$ instead of $\delta V_l$.). In other words, the perturbation $\delta L$ considered is given  by~\cite{Jaramillo:2020tuu}
\begin{eqnarray}\label{delta_L_perturbation_V}
	\delta L=\frac{1}{i}
	\begin{bmatrix}
		0 & 0\\
		\delta q_l/w & 0
	\end{bmatrix}\, .
\end{eqnarray}
At the level of the matrix, the expression $\delta q_l/w$ will be a diagonal matrix. We choose two perturbations of the effective potential at the event horizon and null infinity as follows
\begin{eqnarray}
	\delta q_{l1}&=&A_1\Big\{1-\tanh \Big[H_1(1-x)\Big]\Big\}\, ,\label{delta_ql_1}\label{effective_potential_perturbation_1}\\
	\delta q_{l2}&=&A_2\Big\{1-\tanh \Big[H_2(1+x)\Big]\Big\}\, ,\label{delta_ql_2}\label{effective_potential_perturbation_2}
\end{eqnarray}
respectively. Note that in our setting, the event horizon is located at $x=1$ and the null infinity is located at $x=-1$. The shapes of these two perturbations are shown in Fig.\ref{fig: deltaql}.

In our viewpoints, there are three primary advantages for the considered formulations Eq.(\ref{delta_ql_1}) and Eq.(\ref{delta_ql_2}). First, because the Chebyshev grid is dense at the interval boundary and sparse in the middle, it is more suitable for describing functions that change rapidly at the interval boundary. The functions selected have such characteristics. Second, we notice that each form of perturbation is determined by only two parameters $A$ and $H$, in which $A$ roughly determines the size of the perturbation in term of the energy norm (\ref{energy_norm_sigma}) and $H$ is the shape factor which is used to reflect the degree of concentration of the distribution $\delta q_l$. Only two parameters also allow us to visualize the norm of $\delta L$ intuitively and it is also convenient for us to take parameters along the isoenergetic line. Third, the migration of the spectrum caused by perturbations at the event horizon and the  infinity is what we want to probe, thus Eq.(\ref{delta_ql_1}) and Eq.(\ref{delta_ql_2}) are exactly what we need.

As we know, the fundamental mode $l=m=2$ is the most common mode in astrophysical black holes, which is clearly expected from tensor perturbation and results of numerical relativity~\cite{Franchini:2023eda}, and it holds a very important position in observations~\cite{LIGOScientific:2016lio,LIGOScientific:2020tif,LIGOScientific:2021sio}. Therefore, in order to minimize the variation of parameters but better grasp the research of the (in)stability, we decide to set angular momentum parameter $l$ to $2$ in the following calculations. 

\begin{figure*}
	\centering
	\includegraphics[scale=0.4]{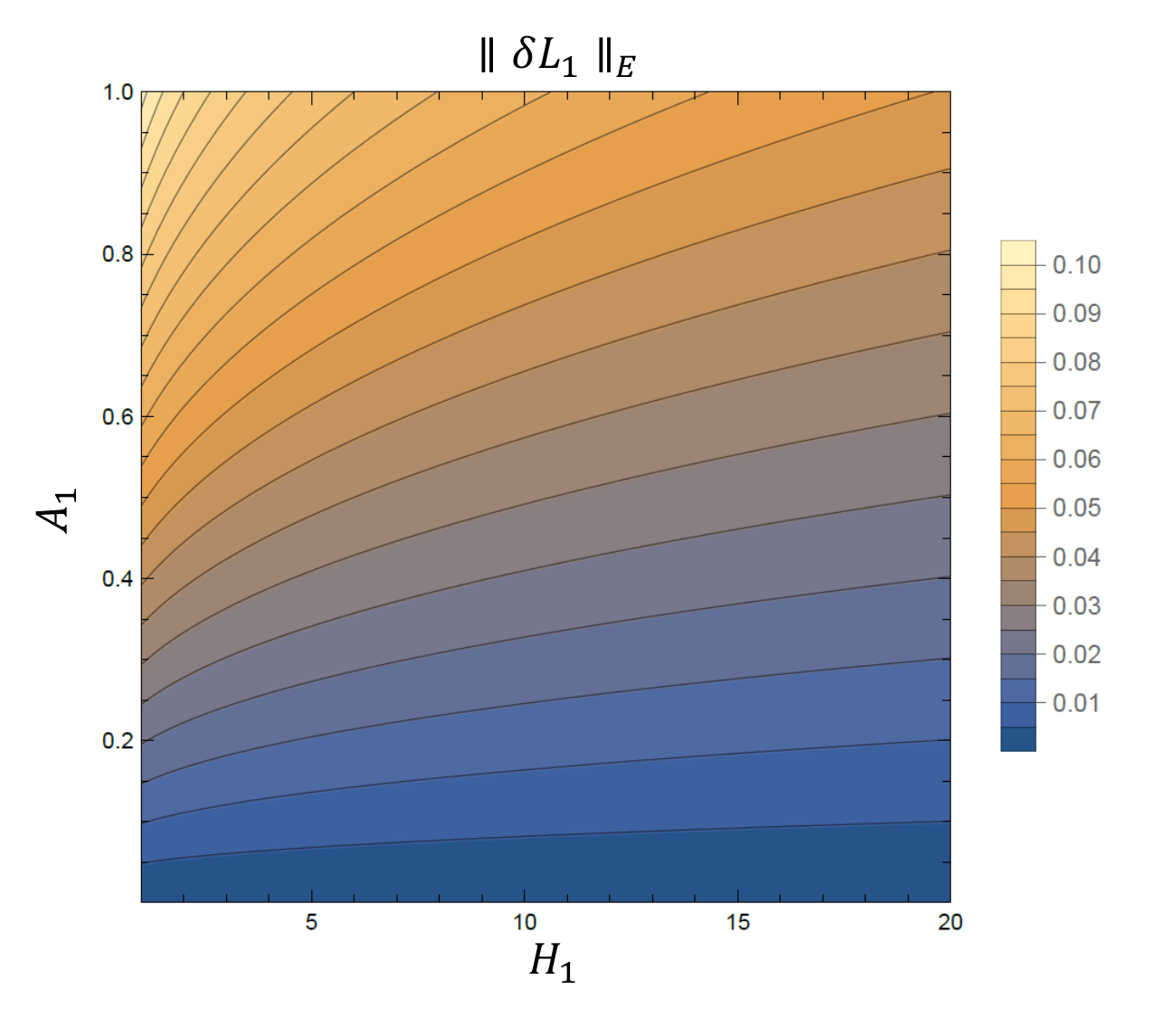}
	\includegraphics[scale=0.4]{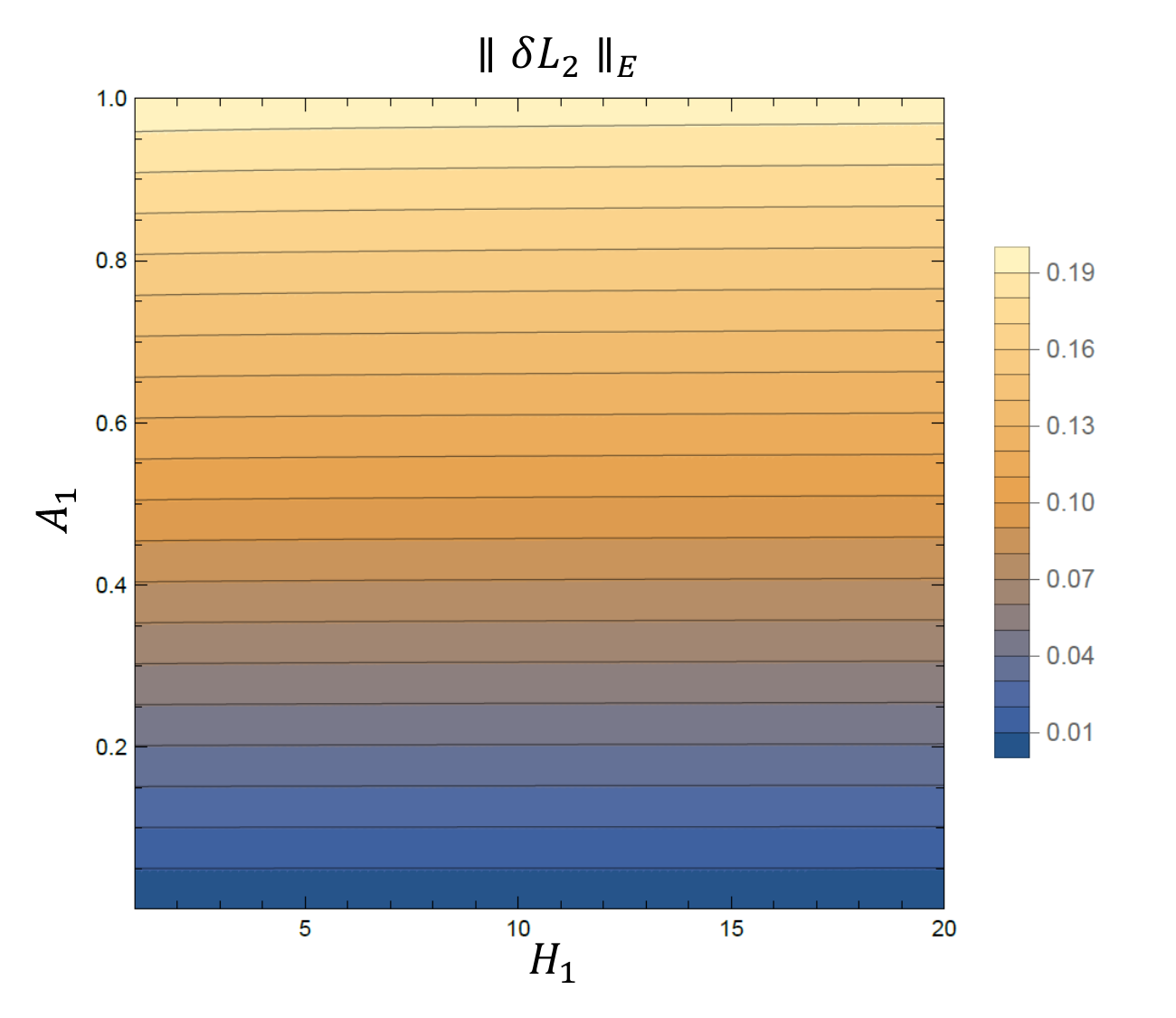}
	\caption{The left panel stands for the energy norm $\lVert\delta L_1\rVert_E$ with perturbation $\delta q_{l1}$ and the right panel stands for the energy norm $\lVert\delta L_2\rVert_E$ with perturbation $\delta q_{l2}$ in the case of $q=0$. In the left panel, there are $20$ contour lines. In the right panel, there are $19$ contour lines.}
	\label{energy_contour_line}
\end{figure*}

For the Schwarzschild black hole case which is the most concerning situation, i.e., $q=0$, the energy contour lines for each perturbation have been shown in Fig.\ref{energy_contour_line}. For comparison purposes, we have set the same parameter ranges with  $H$  and $A$ for both cases. In detail, $H_{\text{min}}=1$, $H_{\text{max}}=20$, $A_{\text{min}}=1/1000$ and $A_{\text{max}}=1$ are considered. We compare the left panel with the right panel and identify the differences between them. First and foremost, the energy contour lines exhibit different behaviors for the two cases. For the case of the perturbation at event horizon, it is found that the energy contour lines show a radial pattern. As $H_1$ decreases, the slopes of the energy contour line increase. However, for the case of the perturbation at null infinity, it can be found that the contour lines are almost parallel, especially for the situation where $A_2$ is relatively small. Second, we find that in terms of the energy norm, the left size is smaller than the right size under the premise of the same parameter ranges.

As a typical example, along the isoenergic line with $\lVert\delta L\rVert_E=0.02$ for the two cases, perturbed characteristic frequencies are derived from the matrix $\textbf{L}+\delta\textbf{L}$, and we observe behaviors of QNM spectra including  the migration and the overtaking instability. The overtaking instability is one of the important manifestations of spectrum instability~\cite{Cheung:2021bol}. The description of migration is as usual provided by Eq.(\ref{delta_omega}) where $q$ has been replaced by $\epsilon$ and $0$ refers to the unperturbed mode. 

\begin{figure*}
	\centering
	\includegraphics[scale=0.45]{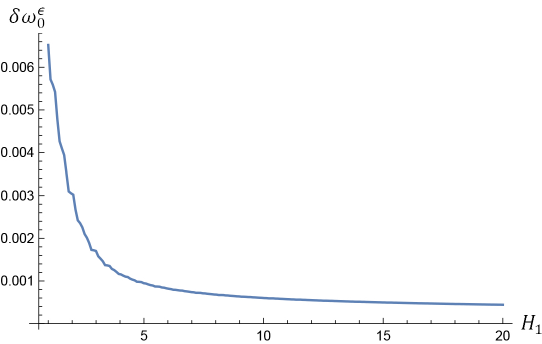}
	\includegraphics[scale=0.45]{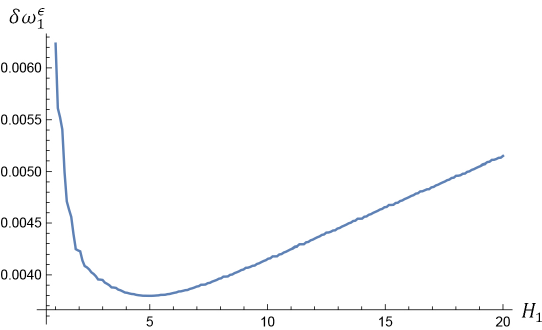}
	\includegraphics[scale=0.45]{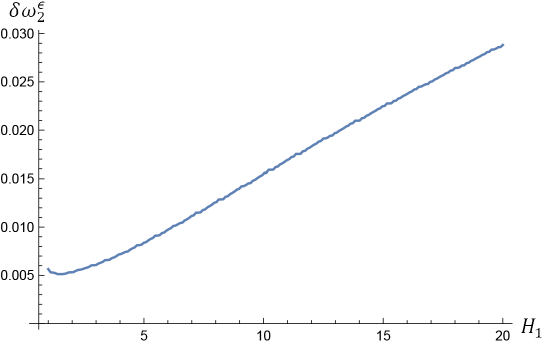}
	\includegraphics[scale=0.45]{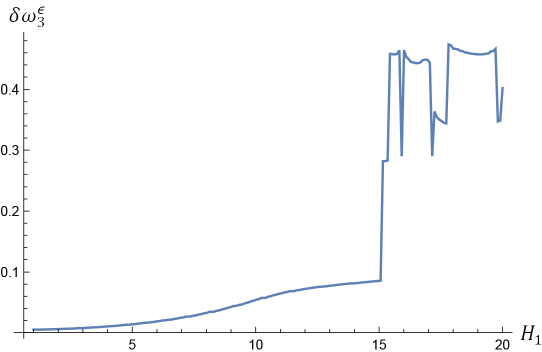}\\
	\includegraphics[scale=0.45]{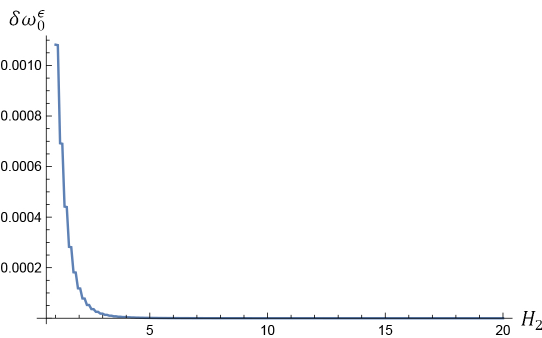}
	\includegraphics[scale=0.45]{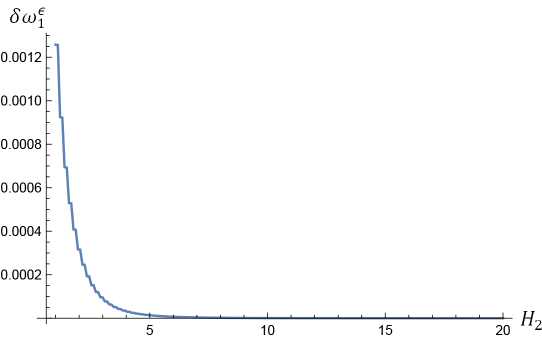}
	\includegraphics[scale=0.45]{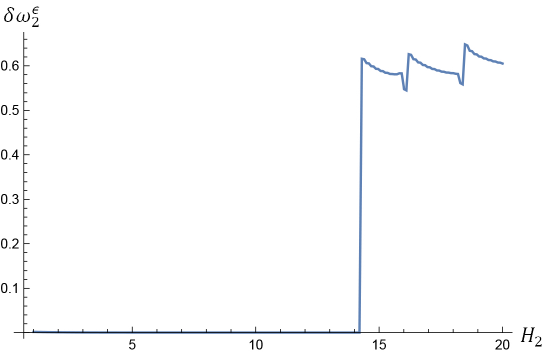}
	\includegraphics[scale=0.45]{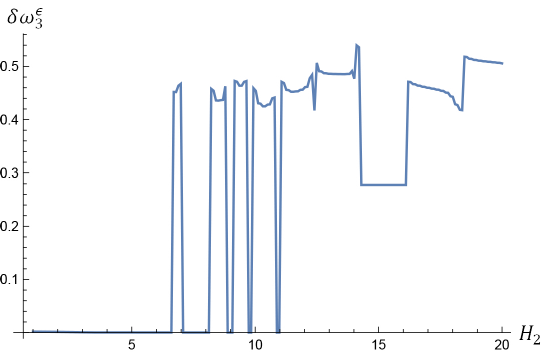}
	\caption{In these eight figures, the energy norms considered are the same, i.e., $\lVert\delta L_1\rVert_E=\lVert\delta L_2\rVert_E=0.02$. The top figures are the migration ratio $\delta\omega_{n}^{\epsilon}$ v.s. the shape factor $H_1$ with $n$ is the overtone number where the perturbation gradually concentrate on the event horizon as $H_1$ increases. The bottom figures are the migration ratios $\delta\omega_{n}^{\epsilon}$ v.s. the shape factor $H_2$ with $n$ is the overtone number where the perturbation gradually concentrate on infinity as $H_2$ increases. The number of grid points considered here is $N=150$.}
	\label{migration_q=0_e=0.02}
\end{figure*}

The numerical results that are shown in Fig.\ref{migration_q=0_e=0.02} under the equal energy norm conditions indicate the following phenomena. First, we find the quantity $\delta \omega^{\epsilon}_n$ has different monotonic behaviors for different overtone numbers $n$. For the top part of Fig.\ref{migration_q=0_e=0.02}, i.e., the case $1$, it is observed that the migration ratio reduces first and then increases for both the first overtone and the second overtone, while the migration ratio of the fundamental mode decreases as increasing $H$. In addition, we can find that the overtaking instability~\cite{Cheung:2021bol} happens when $H_1$ is approximately equal to $15$ for the third overtone. Note that the intuitive manifestation of overtaking is the discontinuity of the image. The reason for occurring the overtaking instability is that the new third mode is replaced by the so-called inner modes~\cite{Jaramillo:2021tmt}. Because the inner modes are very sensitive, when the overtaking happens, the quantity $\delta\omega_3^\epsilon$ oscillates strongly as $H_1$ increases. 

For the bottom part of Fig.\ref{migration_q=0_e=0.02}, i.e., the case $2$, the  behavior of the fundamental mode is similar to the case $1$, and while as $H_2$ increases, it becomes more stable than the case $1$. For the case $2$, it can be found that as the overtone number $n$ increases, the location of $H_2$ decreases in which the overtaking occurs. This also means that as the overtone number $n$ increases, the stability of modes becomes worse. Even for high overtones, overtaking instability can occur many times. What's more, by comparing the top and bottom parts of Fig.\ref{migration_q=0_e=0.02}, the occurrence of overtaking instability caused by perturbations at  infinity is more probable compared to that at the event horizon.

It should be emphasized that we consider the stability of the modes along an isoenergetic line in this work, which is different from~\cite{Cheung:2021bol,Courty:2023rxk} that only keep the amplitude unchanged. Although the amplitude of the perturbation remains constant there, there is no guarantee that the energy carried by the perturbation remains constant. One of the benefits of introducing energy norm (\ref{energy_norm_sigma}) is that it can physically define the energy carried by perturbations. 

From Fig.\ref{fig: deltaql}, it is not difficult to know that the larger $H$ is, the closer the distribution is to the peak distribution. Moreover, Fig.\ref{migration_q=0_e=0.02} indicates that given the energy of the disturbance, using peak distribution can lead to instability of QNMs more efficiently than the average distribution. Visually speaking, perturbations caused by a compact object to the ringdown phase of a black hole are more probable to trigger the overtaking instability of QNM spectrum, as compared to isoenergetic perturbations generated by a gas-like medium. This phenomenon may be valuable for the GWs detection. Additionally, another theoretical estimation~\cite{Gasperin:2021kfv} is provided to indicate that strong peaks in the ``energy distribution" of the perturbation lead to a more pronounced instability compared to the integrated energy distribution, which differs from our visualization scheme. For this point, at a heuristic level, one can get more useful information by referring to such work.

\begin{figure*}
	\centering
	\includegraphics[scale=0.45]{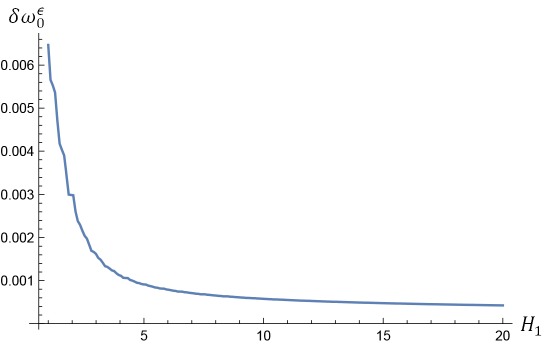}
	\includegraphics[scale=0.45]{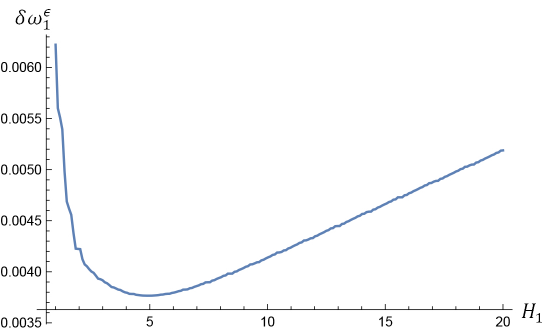}
	\includegraphics[scale=0.45]{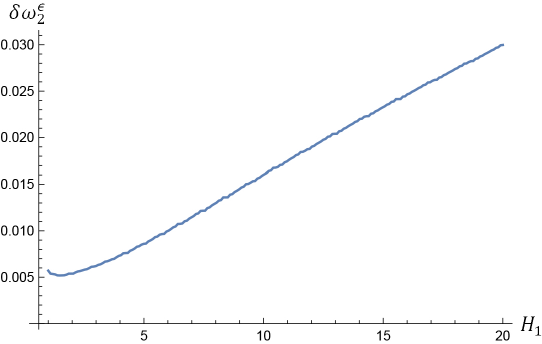}
	\includegraphics[scale=0.45]{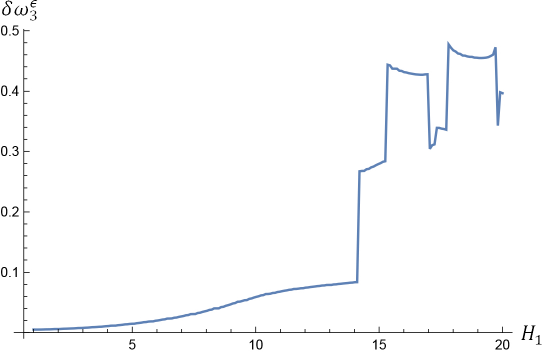}\\
	\includegraphics[scale=0.45]{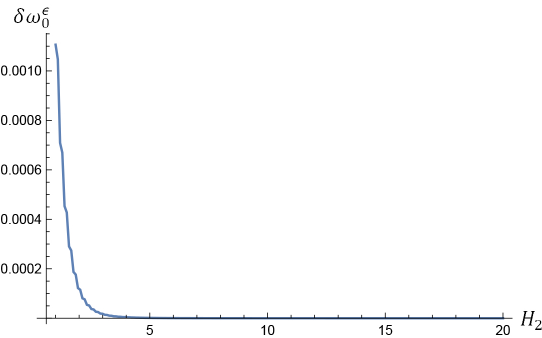}
	\includegraphics[scale=0.45]{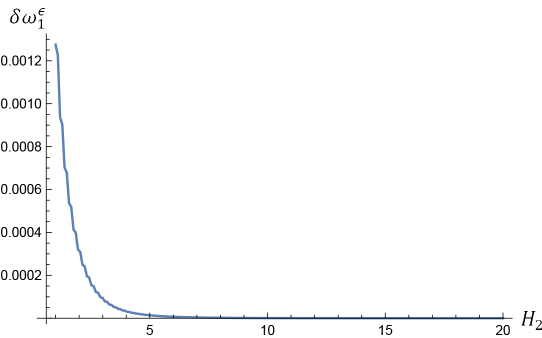}
	\includegraphics[scale=0.45]{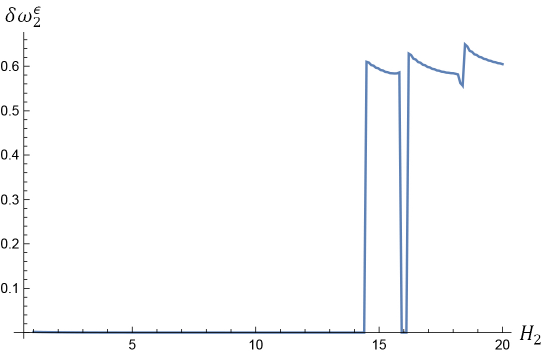}
	\includegraphics[scale=0.45]{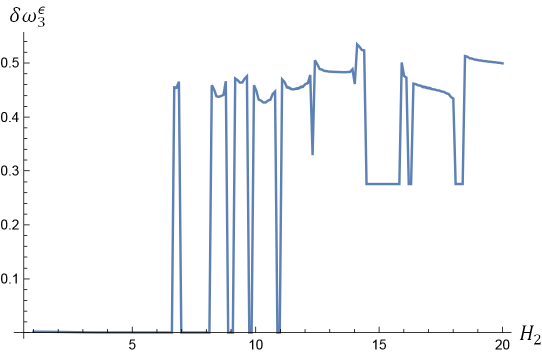}
	\caption{For the case $q=0.5$, the energy norms considered are the same as the case $q=0$, i.e., $\lVert\delta L_1\rVert_E=\lVert\delta L_2\rVert_E=0.02$. These figures are the migration ratio $\delta\omega_{n}^{\epsilon}$ v.s. the shape factor $H$ with $n$ is the overtone number. The tops are for case $1$ and the bottoms are for case $2$. The number of grid points considered here is $N=150$.}
	\label{migration_q=0.5_e=0.02}
\end{figure*}

\begin{figure*}
	\centering
	\includegraphics[scale=0.45]{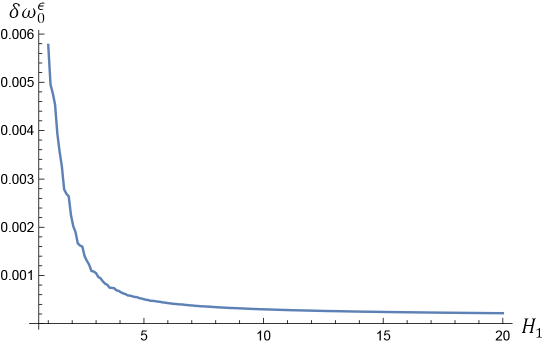}
	\includegraphics[scale=0.45]{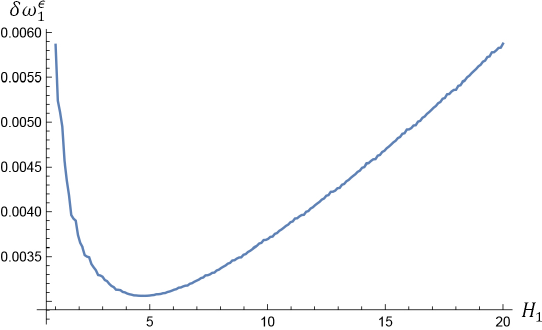}
	\includegraphics[scale=0.45]{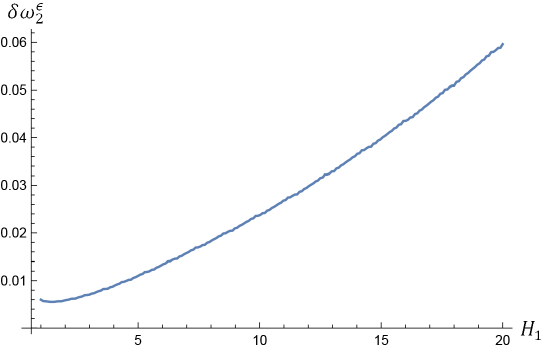}
	\includegraphics[scale=0.45]{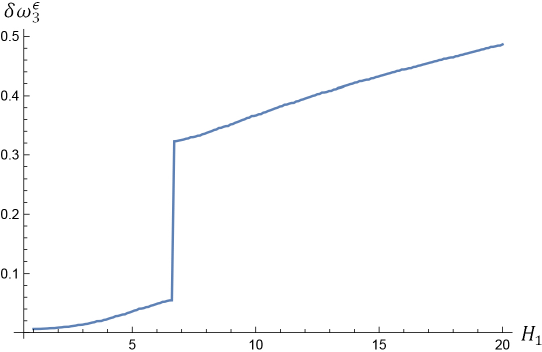}\\
	\includegraphics[scale=0.45]{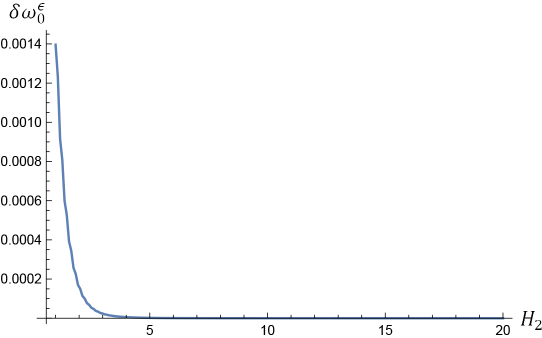}
	\includegraphics[scale=0.45]{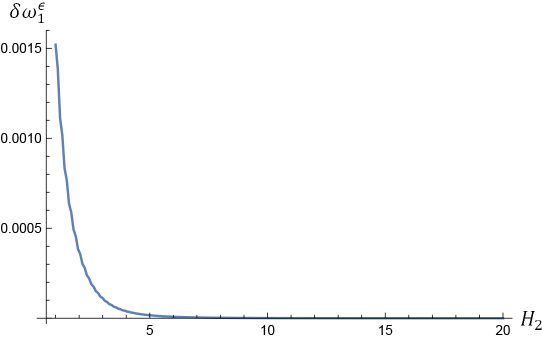}
	\includegraphics[scale=0.45]{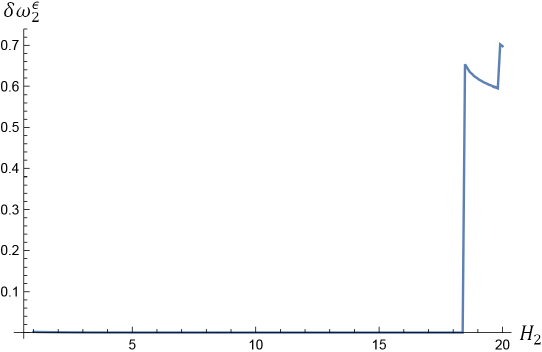}
	\includegraphics[scale=0.45]{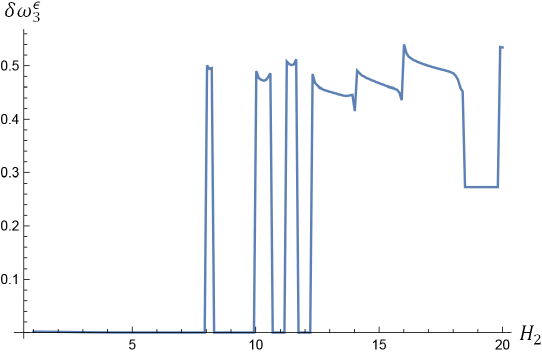}
	\caption{For the case $q=0.8$, the energy norms considered are the same as the case $q=0$, i.e., $\lVert\delta L_1\rVert_E=\lVert\delta L_2\rVert_E=0.02$. These figures are the migration ratios $\delta\omega_{n}^{\epsilon}$ v.s. the shape factor $H$ with $n$ is the overtone number. The tops are for case $1$ and the bottoms are for case $2$. The number of grid points considered here is $N=150$.}
	\label{migration_q=0.8_e=0.02}
\end{figure*}

As for the case $q\neq0$, the qualitative results are similar. For the cases $q=0.5$ and $q=0.8$, we show the migration ratio of modes in Fig.\ref{migration_q=0.5_e=0.02} and Fig.\ref{migration_q=0.8_e=0.02}, respectively. We find that for the case $q=0.5$, even the quantitative results are similar. Only when $q$ is relatively large that corresponds to a small black hole, will the quantitative results have certain differences. It indicates that the spectra of the quantum corrected solution have the almost same (in)stability under such external perturbations [see Eq.(\ref{effective_potential_perturbation_1}) and Eq.(\ref{effective_potential_perturbation_2})] when $q$ is small. Of course, this is only for the migration ratio. Especially noteworthy is that the fundamental mode of all cases is stable irrespective of the location and shape of perturbations. As for behaviours of higher overtones, the peak perturbation at the event horizon will lead to the spectrum instability until the overtaking instability happens, while the peak distribution at the null infinity does not exist such pattern.

Besides, the energy norm of perturbations has always been fixed by $\lVert\delta L\rVert_E=0.02$ in this subsection. The selection of this value has the following reasons. First of all, this value cannot be too small. If the value is small enough, the generation of overtaking instability will occur at higher order overtones from the perspective of the pseudospectrum qualitatively, which is unnecessary for the current situation.  Secondly, for the asymptotically flat spacetime, the branch cut modes, which are not the QNMs, will turn up unlike the dS case or the AdS case. If such norm is large enough, these branch cut modes will also be migrated sharply. At this point, we will face the challenge of determining which are the QNMs and which are not after perturbing. In the bargain,
the inner modes are so unstable that the quantitative performance of the overtaking instability is different for high overtones when we change the number of grid points $N$. But the qualitative phenomenon is the similar.

\section{conclusions and discussion}\label{sec: conclusions}

The impact of quantum corrections on the QNM spectrum of Schwarzschild BHs, as well as the impact of the external environment on the QNM spectrum of such quantum corrected Schwarzschild BHs, which results in the spectrum (in)stability, have been investigated in this paper. The quantum corrected Schwarzschild BH~\cite{Lewandowski:2022zce,Kelly:2020uwj}, as a typical one coming from LQG, is considered in this paper.  Based on this model and previous works about pseudospectrum~\cite{Jaramillo:2020tuu,Destounis:2021lum,Sarkar:2023rhp,Destounis:2023nmb,Arean:2023ejh,Cownden:2023dam,Boyanov:2023qqf}, we study the pseudospectrum and the spectrum (in)stability of such BHs, which will enable us to gain insight of their behaviours.

\begin{figure*}
	\centering
	\includegraphics[scale=0.45]{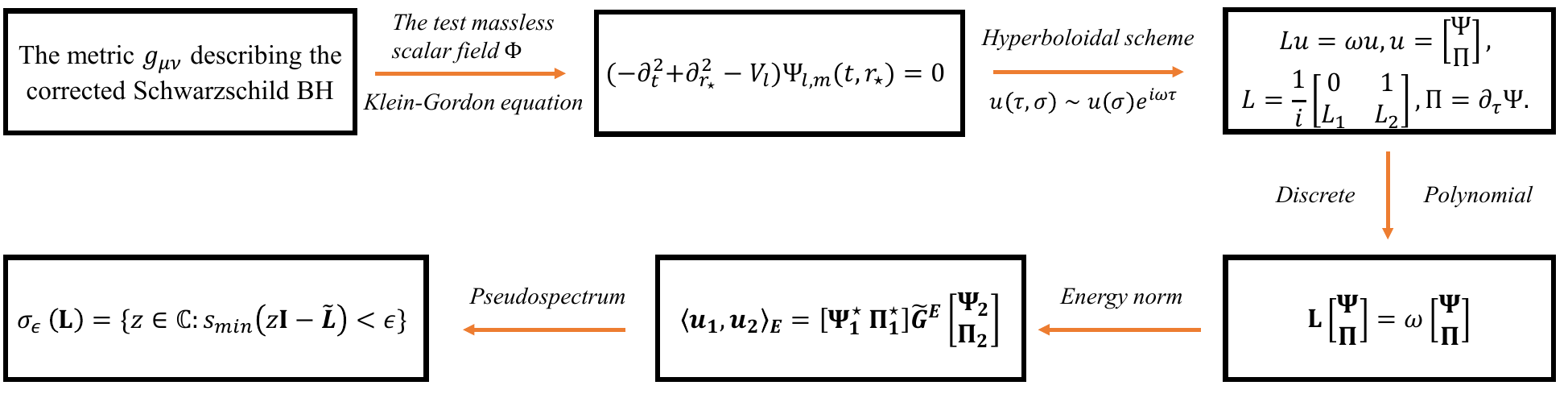}
	\caption{The flow chart to describe the process for dealing with the corrected Schwarzschild BHs and calculating their pseudospectrum.}
	\label{fig:the flow chart}
\end{figure*}

The preparation work for pseudospectrum calculation has been completed in Sec.\ref{sec: set up}. In the first part of Sec.\ref{sec: set up}, we use the hyperboloidal framework to get two partial differential equations of first order in time and second order in space. The advantage of this scheme is that QNM boundary conditions are built into the ``bulk" of the operator $L$. In the second part of Sec.\ref{sec: set up}, a physical energy norm for the QNMs is constructed. The corresponding inner product is also defined by this energy norm. We provide a detailed numerical method for pseudospectrum in the Appendix \ref{app_numerical_approach}. Furthermore, we adopt the invariant subspace projection technique to improve computational speed. What we have done above are the warm up for getting the pseudospectrum. To sum up, this process is universal for dealing with corrected Schwarzschild BHs and calculating their pseudospectrum, there is a flow chart to depict the specific steps in Fig.\ref{fig:the flow chart}.

Our main results are presented in Sec.\ref{sec: stability_of_QNMs}, which has three parts. The first part is the preface of Sec.\ref{sec: stability_of_QNMs}. In the preface, we show the pseudospectrum for the test massless scalar field in the quantum corrected black hole. In the Fig.\ref{fig: pseudospectrum_example}, the ``inverted bridge" structure of pseudospectrum is displayed. Such type of structure reflects the instability of the spectrum  for the QNM problems. The so-called instability means that for small external disturbances, the spectra will undergo significant migrations. Especially for GWs observations, people usually expect the QNM spectrum to be stable. But unfortunately, the instability of the spectrum seems to be an intrinsic property for the QNM problems. Mathematically, the spectral theorem tells us that the spectra of self-adjoint operators are stable. However, such spectral theorem is only valid for the self-adjoint operator. For a QNM problem, the boundary conditions including ingoing and outgoing boundary condition satisfied by some perturbations are dissipative. By consequence, the spectrum problem associated with QNMs is not able to be described by a self-adjoint operator. Therefore, the spectral theorem no longer applies.

The second part is the subsection \ref{sub:A}, where we care about the perturbations caused by the quantum correction in terms of the Schwarzschild black hole background ($q=0$). In addition, we clarify some subtleties about the definition of the perturbation of operator $\delta L$ in this case. To quantitatively describe the (in)stability of QNM spectrum, we defined a so-called migration ratio (\ref{delta_omega}). Using this formula, we describe the migration behaviors for different $l$ and overtone numbers $n$. We find some phenomena as follows. First, as $q$ increases, the migration ratios also increases according to the overall trend of Fig.\ref{fig: migration_q}. Second, as $l$ increases, the migration ratios of different overtones tend to be the same. Third but more important, for $l=0$ and $l=1$, a so-called ``migration ratio instability" will occur between different overtones. 

The third part is the subsection \ref{sub:B}, where we study the spectrum (in)stability of the quantum corrected black hole by adding an additional specific form of perturbation at the event horizon and the null infinity [see Eq.(\ref{effective_potential_perturbation_1}) and Eq.(\ref{effective_potential_perturbation_2})]. Under the premise of controlling the same perturbation energy, we obtain the following conclusions. First, perturbations at infinity are more capable of generating spectrum instability than those at the event horizon. Second, we find that the peak distribution can lead to instability of QNMs more efficiently than the average distribution. 

There are many things that can be further explored associated with such solution. Here, we list some investigations that can be done in the future.
\begin{enumerate}
	\item The black holes in our world are always Kerr type. So in order to better compare the experimental data of GWs, we need to extend this static solution to the rotational case whose related approach is put forward in~\cite{Azreg-Ainou:2014pra}. One obvious thing is to study the shadows and QNMs of this Kerr black hole. What is more important, the pseudospectrum of Kerr type black holes should be taking care of and the hyperboloidal framework for the Kerr spacetime may be helpful~\cite{PanossoMacedo:2019npm}.
	
	\item Beyond QNM spectrum instability, the transient effects associated with the pseudospectrum have been studied in~\cite{Boyanov:2022ark,Jaramillo:2022kuv}. Therefore, we can continue to study the influence of quantum correction $\alpha$ on transient effects.
	
	\item A Weyl's law for QNMs of black holes arouses our interests, since the robustness of the Weyl's law under ultraviolet QNM instability is stated in~\cite{Jaramillo:2022zvf}. Such a law introduces an counting function $N(\omega)$ to estimate of the number of QNMs. Therefore, we will see how the quantum correction $\alpha$ influences such counting function.
\end{enumerate}

\bibliographystyle{unsrt}
\bibliography{reference}


\section*{Acknowledgement}
We are grateful to Long-Yue Li and Xia-Yuan Liu for helpful discussions. This work was supported in part by the National Key R\&D Program of China Grant No.2022YFC2204603. It is also supported by the National Natural Science Foundation of China with grants No.12075232 and No.12247103.

\section*{Conflict of interest}
The authors declare that they have no conflict of interest.



\begin{appendix}
\renewcommand{\thesection}{Appendix}

\section{}
\subsection{\label{app_hyperboloidal_framework} The hyperboloidal framework}
In this appendix, following~\cite{PanossoMacedo:2023qzp}, we will construct the hyperboloidal coordinate $(\tau,\sigma,\theta,\phi)$ via Eq.(\ref{hyperboloidal_coordinate}) in the so-called minimal gauge by imposing 
\begin{eqnarray}
	\rho(\sigma)=\rho_0\, ,
\end{eqnarray}
where $\rho_0$ is a constant. Without loss of generality, we choose the characteristic length scale $\lambda=r_{+}$ in order to fix the black hole event horizon $r_{+}$ at $\sigma=1$. It means that 
$\rho_0=1$. Therefore, Eqs.(\ref{hyperboloidal_coordinate}) reduce to 
\begin{eqnarray}\label{hyperboloidal_coordinate_minimal_gauge}
	\frac{t}{r_{+}}=\tau-H(\sigma)\, ,\quad r=\frac{r_{+}}{\sigma}\, .
\end{eqnarray}
So as to fix $H(\sigma)$ in the minimal gauge, it is convenient to define the dimensionless tortoise coordinate $y(\sigma)$, which is given by
\begin{eqnarray}
	y(\sigma)=\frac{r_{\star}(r(\sigma))}{\lambda}\, .
\end{eqnarray}

In terms of the coordinate $\sigma$, using $r=r_{+}/\sigma$ and Eq.(\ref{q}), the metric function $f(\sigma)$ is expressed as
\begin{eqnarray}
	f(\sigma)&=&(1-\sigma)(1-q^2\sigma)\Big[1+\frac{q^2(q^2+1)\sigma}{1+q^2+q^4}\nonumber\\
	&&+\frac{q^4\sigma^2}{1+q^2+q^4}\Big]\, .
\end{eqnarray}
The derivative of the dimensionless tortoise coordinate is written as
\begin{eqnarray}
	y^{\prime}(\sigma)&=&-\frac{\rho_0}{\sigma^2f(\sigma)}=-\frac{1}{\sigma^2}\Big[1+\frac{1+q^2+q^4+q^6}{1+q^2+q^4}\sigma\nonumber\\
	&&+\mathcal{O}(\sigma^2)\Big]\, .
\end{eqnarray}
Hence, the leading terms contribute with the singular quantities
\begin{eqnarray}
	y_0(\sigma)=\frac{1}{\sigma}-\frac{1+q^2+q^4+q^6}{1+q^2+q^4}\ln\sigma\, .
\end{eqnarray}

\begin{figure*}
	\centering
	\includegraphics[scale=0.7]{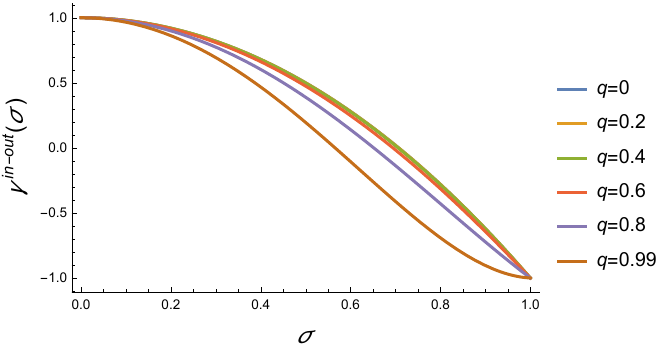}
	\hspace{1cm}
	\includegraphics[scale=0.7]{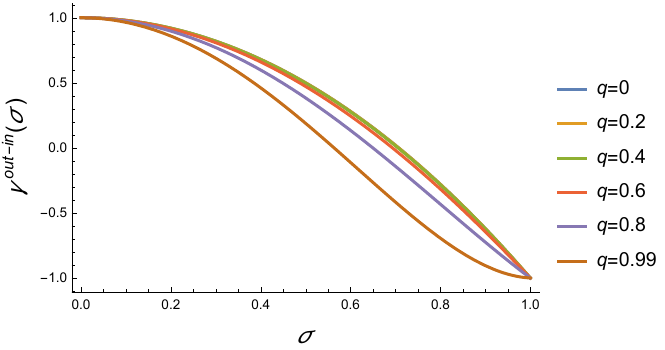}
	\caption{The function $\gamma(\sigma)$ for the quantum corrected black hole with different $q$ in the hyperboloidal slices, where the image on the left is for the function $\gamma^{\text{in-out}}(\sigma)$ and the image on the right is for the function $\gamma^{\text{out-in}}(\sigma)$. The condition $[\gamma(\sigma)]^2<1$ is satisfied when using both the in-out strategy and the out-in strategy within the domain $\sigma\in[0,1]$. It should be noted that these two images look the same, but there is actually a slight difference between them. }
	\label{fig: gamma}
\end{figure*}

At the event horizon $\sigma=1$, the function $K_{+}(\sigma)$ is
\begin{eqnarray}
	K_{+}(\sigma)=(1-q^2\sigma)\Big[1+\frac{q^2(q^2+1)\sigma}{1+q^2+q^4}+\frac{q^4\sigma^2}{1+q^2+q^4}\Big]\, ,
\end{eqnarray}
which induces 
\begin{eqnarray}
	K_{+}(1)=\frac{1+q^2+q^4-3q^6}{1+q^2+q^4}\, .
\end{eqnarray}
Therefore, the function $y_{+}(\sigma)$ is
\begin{eqnarray}
	y_{+}(\sigma)&=&\frac{1}{K_{+}(1)}\ln|\sigma-1|\nonumber\\
	&=&\frac{1+q^2+q^4}{1+q^2+q^4-3q^6}\ln|\sigma-1|\, .
\end{eqnarray}
At the Cauchy horizon $\sigma=1/q^2$, the function $K_{-}(\sigma)$ is
\begin{eqnarray}
	K_{-}(\sigma)=(1-\sigma)\Big[1+\frac{q^2(q^2+1)\sigma}{1+q^2+q^4}+\frac{q^4\sigma^2}{1+q^2+q^4}\Big]\, ,
\end{eqnarray}
which induces
\begin{eqnarray}
	K_{-}\Big(\frac{1}{q^2}\Big)=\frac{-3+q^2+q^4+q^6}{q^2+q^4+q^6}\, .
\end{eqnarray}
Therefore, the function $y_{-}(\sigma)$ is
\begin{eqnarray}
	y_{-}(\sigma)&=&\frac{q^2}{K_{-}(1/q^2)}\ln\Big|\sigma-\frac{1}{q^2}\Big|\nonumber\\
	&=&\frac{q^4+q^6+q^8}{-3+q^2+q^4+q^6}\ln\Big|\sigma-\frac{1}{q^2}\Big|\, .
\end{eqnarray}
The regular piece will then follow from integrating the function $y^{\prime}_{\text{reg}}(\sigma)$~\cite{PanossoMacedo:2023qzp}, in which $y^{\prime}_{\text{reg}}(\sigma)$ is
\begin{eqnarray}
	y^{\prime}_{\text{reg}}(\sigma)&=&y^{\prime}(\sigma)-y^{\prime}_{\text{sing}}(\sigma)=y^{\prime}(\sigma)-y_0^{\prime}(\sigma)-y_{+}^{\prime}(\sigma)-y_{-}^{\prime}(\sigma)\nonumber\\
	&=&\Big\{(q^4+q^2+1) (q^4+2 q^2+3)\nonumber\\
	&&\times(3 q^4+2 q^2+1)[1+q^2+q^4+(q^2+q^4)\sigma+q^4\sigma^2]\Big\}^{-1}\nonumber\\
	&&\times\Big\{q^6\Big[(2+5q^2+4q^4+5q^6+2q^8)\nonumber\\
	&&+(q^2-q^4-q^6+q^8)\sigma\Big]\Big\}\, .
\end{eqnarray}
Unless $q=0$, i.e., the Schwarzschild case, the regular term $y^{\prime}_{\text{reg}}(\sigma)$ is not vanished. This tells us the height function $H^{\text{in-out}}(\sigma)$ for the in-out strategy and the height function $H^{\text{out-in}}(\sigma)$ for the out-in strategy have different results. The expressions of the height function $H^{\text{in-out}}(\sigma)$ and the height function $H^{\text{out-in}}(\sigma)$ are
\begin{eqnarray}
	&&H^{\text{in-out}}(\sigma)\nonumber\\
	&=&\frac{(q^4+q^2+1) \ln (1-\sigma)}{-3 q^6+q^4+q^2+1}-\frac{1}{\sigma}\nonumber\\
	&&+\frac{(q^6+q^4+q^2+1) \ln\sigma}{q^4+q^2+1}+\frac{q^4 (q^4+q^2+1) \log \Big(\sigma -\frac{1}{q^2}\Big)}{q^6+q^4+q^2-3}\nonumber\\
	&&+\frac{q^4}{2 (q^4+q^2+1)(q^4+2 q^2+3) (3 q^4+2 q^2+1)}\nonumber\\
	&&\times\Bigg\{\frac{2 (3 q^8+10 q^6+10 q^4+10 q^2+3) \arctan\Big[\frac{q^2 (2 \sigma +1)+1}{\sqrt{3 q^4+2 q^2+3}}\Big]}{\sqrt{3 q^4+2 q^2+3}}\nonumber\\
	&&+(q^2-1)^2 (q^2+1) \ln \Big[q^4 (\sigma ^2+\sigma +1)+q^2 (\sigma +1)+1\Big]\Bigg\}\, ,\nonumber\\
\end{eqnarray}
and
\begin{eqnarray}
	&&H^{\text{out-in}}(\sigma)\nonumber\\
	&=&\frac{(q^4+q^2+1) \ln (1-\sigma)}{-3 q^6+q^4+q^2+1}-\frac{1}{\sigma}\nonumber\\
	&&+\frac{(q^6+q^4+q^2+1) \ln\sigma}{q^4+q^2+1}+\frac{q^4 (q^4+q^2+1) \log \Big(\sigma -\frac{1}{q^2}\Big)}{q^6+q^4+q^2-3}\nonumber\\
	&&-\frac{q^4}{2 (q^4+q^2+1)(q^4+2 q^2+3) (3 q^4+2 q^2+1)}\nonumber\\
	&&\times\Bigg\{\frac{2 (3 q^8+10 q^6+10 q^4+10 q^2+3) \arctan\Big[\frac{q^2 (2 \sigma +1)+1}{\sqrt{3 q^4+2 q^2+3}}\Big]}{\sqrt{3 q^4+2 q^2+3}}\nonumber\\
	&&+(q^2-1)^2 (q^2+1) \ln \Big[q^4 (\sigma ^2+\sigma +1)+q^2 (\sigma +1)+1\Big]\Bigg\}\, ,\nonumber\\
\end{eqnarray}
respectively. There is a condition, i.e., $[\gamma(\sigma)]^2<1$ that we need to verify, where the so-called boost-function~\cite{Zenginoglu:2011jz} is given by
\begin{eqnarray}
	\gamma(\sigma)=H^{\prime}(\sigma)p(\sigma)\, ,
\end{eqnarray}
with $p(\sigma)=\sigma^2f(\sigma)$. The validation results are shown in Fig.\ref{fig: gamma}.
\subsection{\label{app_numerical_approach} The numerical approach}
In this appendix, we give some necessities for getting the pseudospectrum. First, we employ the Chebyshev spectral method to discretize the differential operator $L$, resulting in its matrix representation $\mathbf{L}$. In order to better apply the formulas related to the Chebyshev polynomial which is naturally defined on the interval $[-1,1]$, it is convenient to map the radial coordinate $\sigma\in[0,1]$ into $x\in[-1,1]$ via a linear transformation
\begin{eqnarray}\label{coordinate_transformation_x_sigma}
	x=2\sigma-1\, ,\quad \sigma=\frac{x+1}{2}\, .
\end{eqnarray}
Therefore, the operator $L_1$ and $L_2$ can be expressed as
\begin{eqnarray}
	L_1(x,q)&=&L_1^0(x,q)+L_1^1(x,q)\frac{\partial}{\partial x}+L_1^2(x,q)\frac{\partial^2}{\partial x^2}\, ,\label{L1_x_q}\\
	L_2(x,q)&=&L_2^0(x,q)+L_2^1(x,q)\frac{\partial}{\partial x}\, ,\label{L2_x_q}
\end{eqnarray}
where the functions $L_1^0(x,q)$, $L_1^1(x,q)$, $L_1^2(x,q)$, $L_2^0(x,q)$ and $L_2^1(x,q)$ are given by
\begin{eqnarray}
	L_1^0(x,q)&=&\Big[q^{18}(x^4+4 x^3+6 x^2-4 x-7)\nonumber\\
	&&+2 q^{16}(x^4+6 x^3+12 x^2-2 x-17)\nonumber\\
	&&+q^{14}(3 x^4+20 x^3+46 x^2-4 x-81)\nonumber\\
	&&+4 q^{12} (x^4+7 x^3+17 x^2-3 x-38)\nonumber\\
	&&+3 q^{10} (x^4+8 x^3+22 x^2-8 x-71)\nonumber\\
	&&+2 q^8(x^4+8 x^3+22 x^2-24 x-119)\nonumber\\
	&&+q^6(x^4+8 x^3+22 x^2-56 x-215)\nonumber\\
	&&-48q^4(x+3)-24q^2(x+3)-8(x+3)\Big]^{-1}\nonumber\\
	&&\times\Big[(q^4+q^2+1)^2\Big(4 l^2(q^4+q^2+1)+4l(q^4+q^2+1)\nonumber\\
	&&-q^6(x^4+4 x^3+6 x^2+2 x-1)\nonumber\\
	&&+2 q^4 (x+1)+2 q^2(x+1)+2(x+1)\Big)\Big]\, ,
\end{eqnarray}
\begin{eqnarray}
	L_1^1(x,q)&=&-\frac{1}{2}\Big[q^{18}(x^4+4 x^3+6 x^2-4 x-7)\nonumber\\
	&&+2 q^{16}(x^4+6 x^3+12 x^2-2 x-17)\nonumber\\
	&&+q^{14}(3 x^4+20 x^3+46 x^2-4 x-81)\nonumber\\
	&&+4 q^{12} (x^4+7 x^3+17 x^2-3 x-38)\nonumber\\
	&&+3 q^{10} (x^4+8 x^3+22 x^2-8 x-71)\nonumber\\
	&&+2 q^8(x^4+8 x^3+22 x^2-24 x-119)\nonumber\\
	&&+q^6(x^4+8 x^3+22 x^2-56 x-215)\nonumber\\
	&&-48q^4(x+3)-24q^2(x+3)-8(x+3)\Big]^{-1}\nonumber\\
	&&\times\Big[(q^4+q^2+1)^2 (x+1) \Big(3 q^6(x^4+4 x^3+6 x^2-3)\nonumber\\
	&&+q^4 (4-12 x)+q^2 (4-12 x)-12 x+4\Big)\Big]\, ,
\end{eqnarray}
\begin{eqnarray}
	L_1^2(x,q)&=&-\frac{1}{4}\Big[q^{18}(x^4+4 x^3+6 x^2-4 x-7)\nonumber\\
	&&+2q^{16}(x^4+6 x^3+12 x^2-2 x-17)\nonumber\\
	&&+q^{14}(3 x^4+20 x^3+46 x^2-4 x-81)\nonumber\\
	&&+4 q^{12} (x^4+7 x^3+17 x^2-3 x-38)\nonumber\\
	&&+3 q^{10} (x^4+8 x^3+22 x^2-8 x-71)\nonumber\\
	&&+2 q^8(x^4+8 x^3+22 x^2-24 x-119)\nonumber\\
	&&+q^6(x^4+8 x^3+22 x^2-56 x-215)\nonumber\\
	&&-48q^4(x+3)-24q^2(x+3)-8(x+3)\Big]^{-1}\nonumber\\
	&&\times\Big[(q^4+q^2+1)^2 (x+1)\Big(q^6 (x+1)^2(x^3+3 x^2+3 x-7)\nonumber\\
	&&-8 q^4 (x^2-1)-8 q^2(x^2-1)-8 x^2+8\Big)\Big]\, ,
\end{eqnarray}
\begin{eqnarray}
	L_2^0(x,q)&=&-\frac{1}{2}\Big[q^{18}(x^4+4 x^3+6 x^2-4 x-7)\nonumber\\
	&&+2 q^{16}(x^4+6 x^3+12 x^2-2 x-17)\nonumber\\
	&&+q^{14}(3 x^4+20 x^3+46 x^2-4 x-81)\nonumber\\
	&&+4 q^{12} (x^4+7 x^3+17 x^2-3 x-38)\nonumber\\
	&&+3 q^{10} (x^4+8 x^3+22 x^2-8 x-71)\nonumber\\
	&&+2 q^8(x^4+8 x^3+22 x^2-24 x-119)\nonumber\\
	&&+q^6(x^4+8 x^3+22 x^2-56 x-215)\nonumber\\
	&&-48q^4(x+3)-24q^2(x+3)-8(x+3)\Big]^{-1}\nonumber\\
	&&\times\Big[(q^4+q^2+1) \Big(q^{12}(5 x^4+20 x^3+30 x^2+4 x-11)\nonumber\\
	&&+q^{10}(5 x^4+28 x^3+54 x^2+12 x-19)\nonumber\\
	&&+q^8(5 x^4+28 x^3+54 x^2-4 x-35)\nonumber\\
	&&+q^6(5 x^4+28 x^3+54 x^2-20 x-51)\nonumber\\
	&&-48 q^4 (x+1)-32 q^2 (x+1)-16 (x+1)\Big)\Big]\, ,
\end{eqnarray}
\begin{eqnarray}
	L_2^1(x,q)&=&-\Big[q^{18}(x^4+4 x^3+6 x^2-4 x-7)\nonumber\\
	&&+2 q^{16}(x^4+6 x^3+12 x^2-2 x-17)\nonumber\\
	&&+q^{14}(3 x^4+20 x^3+46 x^2-4 x-81)\nonumber\\
	&&+4 q^{12} (x^4+7 x^3+17 x^2-3 x-38)\nonumber\\
	&&+3 q^{10} (x^4+8 x^3+22 x^2-8 x-71)\nonumber\\
	&&+2 q^8(x^4+8 x^3+22 x^2-24 x-119)\nonumber\\
	&&+q^6(x^4+8 x^3+22 x^2-56 x-215)\nonumber\\
	&&-48q^4(x+3)-24q^2(x+3)-8(x+3)\Big]^{-1}\nonumber\\
	&&\times\Big[(q^4+q^2+1) \Big(q^{12} (x+1)^2 (x^3+3 x^2+3 x-7)\nonumber\\
	&&+q^{10} (x+1)^2 (x^3+5 x^2+7 x-13)\nonumber\\
	&&+q^8 (x^5+7 x^4+18 x^3-2 x^2-35 x-5)\nonumber\\
	&&+q^6(x^5+7 x^4+18 x^3-10 x^2-51 x+3)\nonumber\\
	&&-24 q^4(x^2+2 x-1)\nonumber\\
	&&-16 q^2(x^2+2 x-1)-8(x^2+2 x-1)\Big)\Big]\, ,
\end{eqnarray}
respectively.

We begin by sampling $N+1$ points from a Chebyshev-Gauss-Labatto grid in which the points are the extrema of Chebyshev polynomial on the interval $[-1,1]$, i.e.,
\begin{eqnarray}\label{Chebyshev_Gauss_Labatto_grid}
	x_{j}=\cos\Big(\frac{j\pi}{N}\Big)\, ,\quad j=0,1,\cdots,N\, ,
\end{eqnarray}
where $x_j\in[-1,1]$. Note that this grid contains two boundary points $x_0=1$ and $x_N=-1$. The differential operators $L$, $L_1$ and $L_2$ will be approximated by using the Chebyshev differentiation matrix $\mathbf{D}_N$. In detail, the first order derivative $\partial/\partial x$ is represented as $\mathbf{D}_N$ and the second order derivative $\partial^2/\partial x^2$ is represented as $\mathbf{D}_N^2$. Moreover, the Chebyshev differentiation matrix $\mathbf{D}_N$ is given by
\begin{eqnarray}\label{Chebyshev_differential_matrix}
	(\mathbf{D}_N)_{00}=\frac{2N^2+1}{6}\, ,\quad (\mathbf{D}_N)_{NN}=-\frac{2N^2+1}{6}\, ,\nonumber\\
	(\mathbf{D}_N)_{jj}=\frac{-x_j}{2(1-x_j^2)}\, ,\quad j=1\,\cdots,N-1\, ,\nonumber\\
	(\mathbf{D}_N)_{ij}=\frac{c_i}{c_j}\frac{(-1)^{i+j}}{(x_i-x_j)}\, ,\quad i\neq j\, ,\quad i,j=0,\cdots,N\, ,
\end{eqnarray}
where
\begin{eqnarray}\label{c_i}
	c_i=\left\{
	\begin{aligned}
		&2&\, ,\quad i=0\, \text{or}\, N\, ,\\
		&1&\, ,\quad \text{otherwise}\, .
	\end{aligned}
	\right. 
\end{eqnarray}
The variables $\Psi$ and $\Pi$ are also discretized as
\begin{eqnarray}
	\mathbf{\Psi}=(\Psi(x_0),\Psi(x_1),\cdots,\Psi(x_{N-1}),\Psi(x_N))^T\, ,
\end{eqnarray}
and
\begin{eqnarray}
	\mathbf{\Pi}=(\Pi(x_0),\Pi(x_1),\cdots,\Pi(x_{N-1}),\Pi(x_N))^T\, ,
\end{eqnarray}
where $\mathbf{\Psi}$ and $\mathbf{\Pi}$ are $N+1$ dimensional vectors.  Therefore, the spectrum problem (\ref{QNM_eigenvalue_problem}) becomes an eigenvalue problem of the matrix, which can be  written as
\begin{eqnarray}\label{QNM_matrix_eigenvalue_problem}
	\mathbf{L}
	\begin{bmatrix}
		\mathbf{\Psi}\\
		\mathbf{\Pi}
	\end{bmatrix}
	=\frac{1}{i}\begin{bmatrix}
		\mathbf{0} & \mathbf{I}\\
		\mathbf{L}_1 & \mathbf{L}_2
	\end{bmatrix}
	\begin{bmatrix}
		\mathbf{\Psi}\\
		\mathbf{\Pi}
	\end{bmatrix}
	=\omega
	\begin{bmatrix}
		\mathbf{\Psi}\\
		\mathbf{\Pi}
	\end{bmatrix}\, .
\end{eqnarray}
Here, the matrix $\mathbf{L}$ on the left is a $2(N+1)\times 2(N+1)$ dimensional matrix and $\mathbf{I}$ is an $(N+1)\times (N+1)$ identity matrix. In terms of details, the matrices $\mathbf{L}_1 $ and $ \mathbf{L}_2$ are explicitly written as
\begin{eqnarray}\label{L1_matrix}
	\mathbf{L}_1&=&\begin{bmatrix}
		L_1^0(x_0,q) & \ & \ \\
		\ & \ddots & \ \\
		\ & \ & L_1^0(x_N,q)
	\end{bmatrix}\mathbf{I}+
	\begin{bmatrix}
		L_1^1(x_0,q) & \ & \ \\
		\ & \ddots & \ \\
		\ & \ & L_1^1(x_N,q)
	\end{bmatrix}\mathbf{D}_N\nonumber\\
	&&+
	\begin{bmatrix}
		L_1^2(x_0,q) & \ & \ \\
		\ & \ddots & \ \\
		\ & \ & L_1^2(x_N,q)
	\end{bmatrix}\mathbf{D}_N^2\, ,
\end{eqnarray}
and
\begin{eqnarray}\label{L2_matrix}
	\mathbf{L}_2=\begin{bmatrix}
		L_2^0(x_0,q) & \ & \ \\
		\ & \ddots & \ \\
		\ & \ & L_2^0(x_N,q)
	\end{bmatrix}\mathbf{I}+
	\begin{bmatrix}
		L_2^1(x_0,q) & \ & \ \\
		\ & \ddots & \ \\
		\ & \ & L_2^1(x_N,q)
	\end{bmatrix}\mathbf{D}_N\, .\nonumber\\
\end{eqnarray}
So far, we have got the matrix version $\mathbf{L}$ of the operator $L$. One can calculate the eigenvalues of $\mathbf{L}$ to obtain QNM frequencies.

Furthermore, in order to calculate the pseudospectrum and study the extent of the perturbations of operator $L$ explicitly, we should deal with the norm (\ref{energy_norm_sigma}) which is a continuous version. It means that we have to use the discretized version of the energy inner product (\ref{energy_inner_product_sigma}). Using the so-called Gram matrix, the continuous version of the inner product (\ref{energy_inner_product_sigma}) can be translated into the matrix quadratic form, i.e.,
\begin{eqnarray}
	\langle u_1,u_2\rangle_E=\begin{bmatrix}
		\mathbf{\Psi}^{\star}_1 & \mathbf{\Pi}^{\star}_1
	\end{bmatrix}
	\tilde{\mathbf{G}}^E
	\begin{bmatrix}
		\mathbf{\Psi}_2\\
		\mathbf{\Pi}_2
	\end{bmatrix}\, ,
\end{eqnarray}
where one can refer to Appendix.\ref{app_Gram_matrix} for the process of obtaining the Gram matrix $\tilde{\mathbf{G}}^E$. Based on the explanation in the previous subsection, we can know that the Gram matrix $\tilde{\mathbf{G}}^E$ is a positive definite Hermitian matrix. Then, the discrete version $\mathbf{L}^{\dagger}$ of the adjoint operator $L^{\dagger}$ is written as
\begin{eqnarray}
	\mathbf{L}^{\dagger}=(\tilde{\mathbf{G}}^E)^{-1}\cdot\mathbf{L}^{\star}\cdot\tilde{\mathbf{G}}^E\, .
\end{eqnarray}
Equipped with the basic components of numerical schemes, our attention turns to the determination of the pseudospectrum. If $\lVert\cdot\rVert=\lVert\cdot\rVert_2$, the norm of a matrix is its largest singular value and the norm of the inverse is the reciprocal of the smallest singular value. In particular~\cite{trefethen2020spectra},
\begin{eqnarray}
	\lVert(z\mathbf{I}-\mathbf{A})^{-1}\rVert_2=[s_{\text{min}}(z\mathbf{I}-\mathbf{A})]^{-1}\, ,
\end{eqnarray}
where $s_{\text{min}}(z\mathbf{I}-\mathbf{A})$ represents the smallest singular value of $z\mathbf{I}-\mathbf{A}$. However, our physical energy norm is not the $2$-norm. It is demanded to establish the connection between a $2$-norm and the energy norm. Thanks to the positive definiteness and Hermiticity for the Gram matrix $\tilde{\mathbf{G}}^E$. There is a useful theorem between the energy norm and the $2$-norm~\cite{trefethen_1999}.

\begin{mythm}
	Given the $2$-norm $\lVert \cdot\rVert_2$ and the energy norm $\lVert \cdot\rVert_E$, for any function $f$, we have
	\begin{eqnarray}\label{2_norm_energy_norm}
		\lVert f(\mathbf{A})\rVert_E=\lVert f(\mathbf{B})\rVert_2\, ,
	\end{eqnarray}
	where the relationship between the matrix $\mathbf{A}$ and the matrix $\mathbf{B}$ satisfies $\mathbf{B}=\mathbf{W}\cdot\mathbf{A}\cdot\mathbf{W}^{-1}$, where the weight matrix $\mathbf{W}$ comes from the Cholesky decomposition of $\tilde{\mathbf{G}}^E$, i.e., $\tilde{\mathbf{G}}^E=\mathbf{W}^{\star}\cdot\mathbf{W}$.
\end{mythm}

At the same time, we identify $\mathbf{A}$ as $\mathbf{L}$ and identify $\mathbf{B}$ as $\tilde{\mathbf{L}}$, and the relation is given by
\begin{eqnarray}
	\tilde{\mathbf{L}}=\mathbf{W}\cdot\mathbf{L}\cdot\mathbf{W}^{-1}\, .
\end{eqnarray}
The above equation tells us that the matrices $\tilde{\textbf{L}}$  and $\textbf{L}$ are connected by a similar transformation matrix $\textbf{W}$. Eq.(\ref{2_norm_energy_norm}) enables us to perform the pseudospectrum computations in the $2$-norm by using the matrix $\tilde{\mathbf{L}}$ rather then the energy norm by using the matrix $\mathbf{L}$. It means that one has
\begin{eqnarray}
	\lVert(z\mathbf{I}-\mathbf{L})^{-1}\rVert_E=\lVert(z\mathbf{I}-\tilde{\mathbf{L}})^{-1}\rVert_2=[s_{\text{min}}(z\mathbf{I}-\tilde{\mathbf{L}})]^{-1}\, .
\end{eqnarray}
From a computational perspective, the definition of the most suitable pseudospectrum for calculation is that~\cite{trefethen2020spectra,Sarkar:2023rhp,trefethen_1999}
\begin{eqnarray}
	\sigma_{\epsilon}(\mathbf{L})=\{z\in\mathbb{C}:s_{\text{min}}(z\mathbf{I}-\tilde{\mathbf{L}})<\epsilon\}\, .
\end{eqnarray}
So the pseudospectrum of $\mathbf{L}$ constructs the sets in the $z$-plane bounded by contour curves of the function $s_{\text{min}}(z\mathbf{I}-\tilde{\mathbf{L}})$. $z$-plane is nothing but the frequency $\omega$-plane for our QNM problem. 

The basic algorithm is the singular value decomposition (SVD) method. We will select the minimum value among the results of the singular value for the matrix $z\mathbf{I}-\tilde{\mathbf{L}}$ at each point $z\in\mathbb{C}$. The basic algorithm described is simple but inefficient when the matrix $z\mathbf{I}-\tilde{\mathbf{L}}$ is large. Therefore, in Appendix.\ref{app_subspace_method}, invariant subspace method can help us to reduce the calculation size and maintain a certain level of accuracy simultaneously. Finally, the calculation of pseudospectrum is very suitable for parallelization, as the calculation of the minimum singular value at each point on the two-dimensional grid ($z$-plane) is independent of each other.

\subsection{\label{app_Gram_matrix} The Gram matrix $\tilde{\mathbf{G}}^E$}
In this appendix, we will construct the Gram matrix following~\cite{Jaramillo:2020tuu,Sheikh:2022cud}. 
For the convenience of numerical applications, we are supposed to transform the integrals (\ref{energy_inner_product_sigma}) in terms of $\sigma$ into the integrals in terms of $x$ by using Eq.(\ref{coordinate_transformation_x_sigma}), i.e.,
\begin{eqnarray}\label{energy_inner_product_x}
	\langle u_1,u_2\rangle_E&=&\frac14\int_{-1}^1\Big[w\Big(\frac{x+1}{2}\Big)\Pi^{\star}_1\Pi_2+4\times p\Big(\frac{x+1}{2}\Big)\partial_x\Psi^{\star}_1\partial_x\Psi_2\nonumber\\
	&&+q_l\Big(\frac{x+1}{2}\Big)\Psi^{\star}_1\Psi_2\Big]\mathrm{d}x\, ,
\end{eqnarray}
where we have 
\begin{eqnarray}
	w\Big(\frac{x+1}{2}\Big)&=&-\frac{1}{4(q^4+q^2+1)^3}\Big[q^{18}(x^4+4 x^3+6 x^2-4 x-7)\nonumber\\
	&&+2 q^{16} (x^4+6 x^3+12 x^2-2 x-17)\nonumber\\
	&&+q^{14}(3 x^4+20 x^3+46 x^2-4 x-81)\nonumber\\
	&&+4 q^{12}(x^4+7 x^3+17 x^2-3 x-38)\nonumber\\
	&&+3 q^{10}(x^4+8 x^3+22 x^2-8 x-71)\nonumber\\
	&&+2 q^8 (x^4+8 x^3+22 x^2-24 x-119)\nonumber\\
	&&+q^6(x^4+8 x^3+22 x^2-56 x-215)\nonumber\\
	&&-48q^4(x+3)-24q^2(x+3)-8(x+3)\Big]\, ,
\end{eqnarray}
\begin{eqnarray}
	p\Big(\frac{x+1}{2}\Big)&=&\frac{1}{64 (q^4+q^2+1)}\Big\{(x-1) (x+1)^2 [q^2 (x+1)-2]\nonumber\\
	&&\times[q^4 (x^2+4 x+7)+2 q^2 (x+3)+4]\Big\}\, ,
\end{eqnarray}
\begin{eqnarray}
	q_l\Big(\frac{x+1}{2}\Big)&=&\frac{1}{4(q^4+q^2+1)}\Big[4 l^2 (q^4+q^2+1)+4 l (q^4+q^2+1)\nonumber\\
	&&+q^6 (-x^4-4x^3-6x^2-2x+1)\nonumber\\
	&&+2 q^4 (x+1)+2 q^2 (x+1)+2 (x+1)\Big]\, .
\end{eqnarray}
In other words, we should consider the quadrature approximation of the integrals (\ref{energy_inner_product_x}). First, we generally take into account an integral whose upper and lower limits are $1$, $-1$ with definition as
\begin{eqnarray}\label{I_mu}
	I_{\mu}(f,g)\equiv\int_{-1}^1f(x)g(x)\mu(x)\mathrm{d}x\, ,
\end{eqnarray}
where $\mu(x)$ is called the weight function. Using the Chebyshev-Gauss-Labatto grid (\ref{Chebyshev_Gauss_Labatto_grid}), we obtain the quadrature approximation of the above integral (\ref{I_mu}) as follows
\begin{eqnarray}\label{I_mu_discrete}
	I_{\mu}(f,g)\approx I^N_{\mu}(f,g)=\mathbf{f}^T_{N}\cdot \mathbf{C}^N_{\mu}\cdot \mathbf{g}_{N}\, ,
\end{eqnarray}
in which $\cdot$ is the usual matrix multiplication and $\mathbf{f}^T_N=(f(x_0),\cdots,f(x_N))$, $\mathbf{g}_N=(g(x_0),\cdots,g(x_N))^T$ are derived by substituting the grid (\ref{Chebyshev_Gauss_Labatto_grid}) into the functions $f(x)$ and $g(x)$. The key ingredient is the expression of the $(N+1)\times(N+1)$ intermediate matrix $\mathbf{C}_{\mu}^N$ which is a diagonal matrix in the Eq.(\ref{I_mu_discrete}). Its diagonal elements are
\begin{eqnarray}
	(\mathbf{C}^N_{\mu})_{ii}&=&\frac{2\mu(x_i)}{c_iN}\Big[1-\sum_{k=1}^{\lfloor \frac{N}{2} \rfloor}T_{2k}(x_i)\frac{2-\delta_{2k,N}}{4k^2-1}\Big]\, ,\nonumber\\
	&&i=0,1,\cdots,N-1,N\, ,
\end{eqnarray}
where $\lfloor a \rfloor$ is the floor function, i.e., the largest integer that is less than or equal to $a$, and $T_{k}$ is the Chebyshev polynomial of order $k$. Next, using Eq.(\ref{I_mu_discrete}), we write the discrete version of the inner product (\ref{energy_inner_product_x}) as
\begin{eqnarray}\label{discrete_energy_inner_product_x}
	\langle u_1,u_2\rangle_E&=&\Big\langle\begin{bmatrix}
		\Psi_1\\
		\Pi_1
	\end{bmatrix}\, ,
	\begin{bmatrix}
		\Psi_2\\
		\Pi_2
	\end{bmatrix}\Big\rangle_E\nonumber\\
	&=&\frac{1}{4}\mathbf{\Pi}_1^{\star}\cdot \mathbf{C}^N_{w}\cdot \mathbf{\Pi}_2+(\mathbf{D}_N\cdot\mathbf{\Psi}_1)^{\star}\cdot \mathbf{C}_{p}^N\cdot(\mathbf{D}_N\cdot\mathbf{\Psi}_2)\nonumber\\
	&&+\frac14\mathbf{\Psi}_1^{\star}\cdot \mathbf{C}_{q_l}^N\cdot\mathbf{\Psi}_2\nonumber\\
	&=&\begin{bmatrix}
		\mathbf{\Psi}^{\star}_1 & \mathbf{\Pi}^{\star}_1
	\end{bmatrix}
	\begin{bmatrix}
		\frac{1}{4}\mathbf{C}^N_{q_l}+\mathbf{D}_N^{\star}\cdot \mathbf{C}^N_p\cdot \mathbf{D}_N & \mathbf{0}\\
		\mathbf{0} & \frac{1}{4}\mathbf{C}^N_w
	\end{bmatrix}
	\begin{bmatrix}
		\mathbf{\Psi}_2\\
		\mathbf{\Pi}_2
	\end{bmatrix}\, ,\nonumber\\
\end{eqnarray}
where $\mathbf{D}_N$ is given by Eq.(\ref{Chebyshev_differential_matrix}) and $\star$ stands for the Hermitian conjugation of the matrix. Eq.(\ref{discrete_energy_inner_product_x}) allows us to define the Gram matrix $\mathbf{G}^E$ for the discretized version of the energy scalar product (\ref{energy_inner_product_x}), where $\mathbf{G}^E$ is given by
\begin{eqnarray}
	\mathbf{G}^E\equiv\begin{bmatrix}
		\mathbf{G}^E_1 & \mathbf{0}\\
		\mathbf{0} & \mathbf{G}_2^E
	\end{bmatrix}
	=\begin{bmatrix}
		\frac{1}{4}\mathbf{C}^N_{q_l}+\mathbf{D}_N^{\star}\cdot \mathbf{C}^N_p\cdot \mathbf{D}_N & \mathbf{0}\\
		\mathbf{0} & \frac{1}{4}\mathbf{C}^N_w
	\end{bmatrix}\, .
\end{eqnarray}
This is a $2(N+1)\times 2(N+1)$ dimensional matrix, in which $\mathbf{G}_1^E$ and $\mathbf{G}_2^E$ read as
\begin{eqnarray}
	\mathbf{G}_1^E&=&\frac{1}{4}\mathbf{C}^N_{q_l}+\mathbf{D}_N^{\star}\cdot \mathbf{C}^N_p\cdot \mathbf{D}_N\, ,\nonumber\\
	\mathbf{G}_2^E&=&\frac{1}{4}\mathbf{C}^N_w\, .
\end{eqnarray}

With some accuracy issue, the Gram matrix $\mathbf{G}^E$ is not the one we used to calculate the norm. Instead, we should modify the Gram matrix $\mathbf{G}^E$ by incorporating the interpolation strategy. If we have two sets of Chebyshev-Gauss-Lobatto grids denoted as $\{x_i\}^N_{i=0}$ and $\{\bar{x}_i\}^{\bar{N}}_{i=0}$ which are associated interpolant vectors $f_{N}(x_i)$ and $f_{\bar{N}}(\bar{x}_i)$, then the relation between $f_{N}(x_i)$ and $f_{\bar{N}}(\bar{x}_i)$ is given by
\begin{eqnarray}
	f_{\bar{N}}(\bar{x}_i)=\sum_{i=0}^N\mathbf{P}_{\bar{i}i}f_N(x_i)\, ,
\end{eqnarray}
where
\begin{eqnarray}
	\mathbf{P}_{\bar{i}i}&=&\frac{1}{c_iN}\Big[1+\sum_{j=1}^N(2-\delta_{j,N})T_j(\bar{x}_i)T_j(x_i)\Big]\, ,\nonumber\\
	&&\bar{i}=0,\cdots,\bar{N}\quad \text{and}\quad i=0,\cdots,N\, ,
\end{eqnarray}
with the grid $\{\bar{x}_i\}^{\bar{N}}_{i=0}$ is defined as
\begin{eqnarray}\label{new_Chebyshev_Gauss_Labatto_grid}
	\bar{x}_{j}=\cos\Big(\frac{\bar{j}\pi}{\bar{N}}\Big)\, ,\quad \bar{j}=0,1,\cdots,\bar{N}\, .
\end{eqnarray}
It should be noted that $\mathbf{P}$ is a matrix with size $(\bar{N}+1)\times (N+1)$ and $c_i$ is defined by Eq.(\ref{c_i}). In fact, when $\bar{N}=N$, $\mathbf{P}$ is the identity matrix with size $(N+1)\times (N+1)$.

Now, for a fixed $N$, the discrete integration (\ref{I_mu_discrete}) in terms of a higher resolution $\bar{N}=2N+1$ is considered here. Generally speaking, $\bar{N}$ can take any value greater than $N$. We then interpolate the expression back to the original resolution $N$. That is to say, defining 
\begin{eqnarray}
	\mathcal{I}_\mu^N(f,g):=I_\mu^{\bar{N}}(f,g)\, ,
\end{eqnarray}
we get the grid-interpolated new discrete integration as follows
\begin{eqnarray}
	\mathcal{I}_\mu^N(f,g)=\mathbf{f}^T_N\cdot\tilde{\mathbf{C}}^N_\mu\cdot \mathbf{g}_N\, ,
\end{eqnarray}
where the matrix $\tilde{\mathbf{C}}^N_\mu$ is obtained by
\begin{eqnarray}
	\tilde{\mathbf{C}}^N_\mu=\mathbf{P}^T\cdot \mathbf{C}^{\bar{N}}_\mu\cdot \mathbf{P}\, .
\end{eqnarray}
In other words, the components of $\tilde{\mathbf{C}}^N_\mu$ are expressed as
\begin{eqnarray}
	(\tilde{\mathbf{C}}^N_\mu)_{ij}=\sum^{\bar{N}}_{\bar{i}=0}\sum^{\bar{N}}_{\bar{j}=0}(\mathbf{P}^T)_{i\bar{i}}(\mathbf{C}^{\bar{N}}_{\mu})_{\bar{i}\bar{j}}\mathbf{P}_{\bar{j}j}\, .
\end{eqnarray}
Therefore, the new Gram matrix allows us to calculate the inner product (\ref{discrete_energy_inner_product_x}) via
\begin{eqnarray}\label{discrete_energy_inner_product_x_new}
	\langle u_1,u_2\rangle_E&=&\begin{bmatrix}
		\mathbf{\Psi}^{\star}_1 & \mathbf{\Pi}^{\star}_1
	\end{bmatrix}
	\tilde{\mathbf{G}}^E
	\begin{bmatrix}
		\mathbf{\Psi}_2\\
		\mathbf{\Pi}_2
	\end{bmatrix}
	\nonumber\\
	&=&\begin{bmatrix}
		\mathbf{\Psi}^{\star}_1 & \mathbf{\Pi}^{\star}_1
	\end{bmatrix}
	\begin{bmatrix}
		\tilde{\mathbf{G}}_1^E & \mathbf{0}\\
		\mathbf{0} & \tilde{\mathbf{G}}_2^E
	\end{bmatrix}
	\begin{bmatrix}
		\mathbf{\Psi}_2\\
		\mathbf{\Pi}_2
	\end{bmatrix}\, ,
\end{eqnarray}
in which we grid interpolate the Gram matrix $\mathbf{G}^E$ as follows
\begin{eqnarray}
	\tilde{\mathbf{G}}_1^E=\mathbf{P}^T\cdot \mathbf{G}_1^E\cdot\mathbf{P}\, ,\nonumber\\
	\tilde{\mathbf{G}}_2^E=\mathbf{P}^T\cdot \mathbf{G}_2^E\cdot\mathbf{P}\, .
\end{eqnarray}
It should be noted that, on the one hand, we still discretize the differential operator $L$ in a Chebyshev grid with $N+1$ points [see Eq.(\ref{Chebyshev_Gauss_Labatto_grid})] and get a matrix $\mathbf{L}$ with size $2(N+1)\times2(N+1)$ whose structure is given by Eq.(\ref{QNM_matrix_eigenvalue_problem}). On the other hand, we discretize in a finer Chebyshev grid with $\bar{N}+1=2(N+1)$ points [see Eq.(\ref{new_Chebyshev_Gauss_Labatto_grid})] to obtain a Gram matrix $\mathbf{G}^E=\text{diag}(\mathbf{G}^E_1,\mathbf{G}^E_2)$ with size $4(N+1)\times4(N+1)$. Then, we interpolate to the coarser Chebyshev grid [see Eq.(\ref{Chebyshev_Gauss_Labatto_grid})] to get a final Gram matrix $\tilde{\mathbf{G}}^E=\text{diag}(\tilde{\mathbf{G}}^E_1,\tilde{\mathbf{G}}^E_2)$ with size $2(N+1)\times2(N+1)$. Naturally and necessarily, the sizes of $\mathbf{L}$  and $\tilde{\mathbf{G}}^E$ are consistent.

\subsection{\label{app_subspace_method} The subspace method for the pseudospectrum}
In this appendix, following the monograph~\cite{trefethen2020spectra}, we introduce an important technique---invariant subspace method that reduces the matrix to a fraction of its original size by orthogonally projecting the matrix onto an appropriate low-dimensional subspace so as to improve computational efficiency. Strictly speaking, this subspace method is only an approximate method for calculating pseudospectrum. However, within the region of our interest, this approximation is already accurate enough to illustrate the (in)stability problem of the QNMs.

Assuming that the object for calculating pseudospectrum is matrix $\mathbf{B}$ whose dimension is $N$ and we take the projection subspace $\mathcal{U}\subset\mathbb{C}^N$ to be the invariant subspace, associated with $\mathbf{B}$. If $\mathbf{U}\in\mathbb{C}^{N\times p}$ is a matrix whose columns form an orthonromal basis for $\mathcal{U}$, then the eigenvalues of the projected matrix $\mathbf{U}^{\star}\mathbf{B}\mathbf{U}$ are simply satisfy
\begin{eqnarray}
	\sigma(\mathbf{U}^{\star}\mathbf{B}\mathbf{U}) \subseteq \sigma(\mathbf{B})\, .
\end{eqnarray}
Therefore, we have the following theorem involving the pseudospectrum of the projected matrix $\mathbf{U}^{\star}\mathbf{B}\mathbf{U}$ and the original matrix $\mathbf{B}$ (One can find this theorem in Chapter $40$ of the monograph~\cite{trefethen2020spectra}.).

\begin{mythm}
	If {\rm range($\mathbf{U}$)} is an invariant subspace of $\mathbf{B}$, then
	\begin{eqnarray}
		\sigma_{\epsilon}(\mathbf{U}^{\star}\mathbf{B}\mathbf{U})\subseteq\sigma_{\epsilon}(\mathbf{B}).
	\end{eqnarray}
\end{mythm}

How do we obtain the subspace $\mathcal{U}$ and its representation $\mathbf{U}$? One can compute the Schur factorization of $\mathbf{B}$, and reorder this decomposition to put the $p$ desired eigenvalues in the first $p$ positions on the diagonal of the Schur factor $\mathbf{T}$. The unitary factor in the Schur decomosition is converted into the form $[\mathbf{U}\, \hat{\mathbf{U}}]$ by such reordering, where $\mathbf{U}$ has orthonormal columns spanning $\mathcal{U}$ as we required. In mathematical terms, the Schur factorization takes the form
\begin{eqnarray}
	\mathbf{T}=[\mathbf{U}\, \hat{\mathbf{U}}]^{\star}\mathbf{B}[\mathbf{U}\, \hat{\mathbf{U}}]=
	\begin{bmatrix}
		\mathbf{U}^{\star}\mathbf{B}\mathbf{U} & \mathbf{U}^{\star}\mathbf{B}\hat{\mathbf{U}}\\
		\hat{\mathbf{U}}^{\star}\mathbf{B}\mathbf{U} & \hat{\mathbf{U}}^{\star}\mathbf{B}\hat{\mathbf{U}}
	\end{bmatrix}=
	\begin{bmatrix}
		\mathbf{T}_{11} & \mathbf{T}_{12}\\
		\mathbf{0} & \mathbf{T}_{22}
	\end{bmatrix}\, ,
\end{eqnarray}
where $\mathbf{T}_{11}$ and $\mathbf{T}_{22}$ are upper triangular matrices, with $\mathbf{T}_{11}$ equals to the desired compression $\mathbf{U}^{\star}\mathbf{B}\mathbf{U}$.

Based on the frequency distribution characteristics of QNMs, we choose the subspace $\mathcal{U}$ that is generated by the eigenvectors corresponding to all eigenvalues whose imaginary parts are smaller than a given value. This upper bound denoted as $I_{\text{max}}$ is choosen adjustably. Intuitively speaking, the larger the selected upper bound, the more accurate the pseudospectrum obtained, but the more time it takes to calculate.

\end{appendix}
\end{multicols}

\end{document}